\documentclass[manuscript]{aastex}
\makeindex
\makeatletter

\usepackage{natbib}
\usepackage{booktabs}
\newcommand{\ra}[1]{\renewcommand{\arraystretch}{#1}}
\usepackage{color}

\newcommand{\veczhat}{{\mathbf {\hat z}}}

\def \bvec{\bf{B}}
\def \vvec{\bf{V}}
\def \evec{\bf{E}}

\def \szpot{{S$_{z,p}$}}

\def \szfree{{S$_{z,f}$}}
\def \szpot{{S$_{z,p}$}}
\def \sztot{{S$_{z}$}}

\newcommand{\be}{\begin{equation}}
\newcommand{\ee}{\end{equation}}

\shorttitle{Photospheric E-Fields \& Energy Fluxes in AR 11158 
} \shortauthors{Kazachenko et. al}
\def\Rsun{\ifmmode{R_\odot}\else{R$_\odot$}\fi}

\begin{document}

\title{
Photospheric Electric Fields and Energy Fluxes in the Eruptive Active Region NOAA 11158 
}

\author{Maria D. Kazachenko\altaffilmark{1}, George H. Fisher\altaffilmark{1}, Brian T. Welsch\altaffilmark{1}, Yang Liu\altaffilmark{2}, Xudong Sun\altaffilmark{2}}

\altaffiltext{1}{Space Sciences Laboratory, UC Berkeley, CA 94720, USA}
\altaffiltext{2}{W. W. Hansen Experimental Physics Laboratory, Stanford University, Stanford, CA 94305, USA}

\email{kazachenko@ssl.berkeley.edu}

\begin{abstract}
How much electromagnetic energy crosses the photosphere in evolving solar active regions?  With the advent of high-cadence vector magnetic field observations, addressing this fundamental question has become tractable. In this paper, we apply the ``PTD-Doppler-FLCT-Ideal'' (\texttt{PDFI}) electric field inversion technique of \citet{Kazachenko2014a} to a 6-day HMI/SDO vector magnetogram and Doppler velocity sequence, to find the electric field and Poynting flux evolution in active region NOAA 11158, which produced an X2.2 flare early on 2011 February 15.  We find photospheric electric fields ranging up to $2$ V/cm. The Poynting fluxes range from $[-0.6$ to $2.3]\times10^{10}$ ergs$\cdot$cm$^{-2}$s$^{-1}$, mostly positive, with the largest contribution to the energy budget in the range of $[10^9$--$10^{10}]$ ergs$\cdot$cm$^{-2}$s$^{-1}$. Integrating the instantaneous energy flux over space and time, we find that the total magnetic energy accumulated above the photosphere from the initial emergence to the moment before the X2.2 flare to be $E=10.6\times10^{32}$ ergs, which is partitioned as $2.0$ and $8.6\times10^{32}$ ergs, respectively, between free and potential energies. Those estimates are consistent with estimates from preflare non-linear force-free field (NLFFF) extrapolations and the Minimum Current Corona estimates (MCC), in spite of our very different approach. This study of  photospheric electric fields demonstrates the potential of the \texttt{PDFI} approach for estimating Poynting fluxes and opens the door to more quantitative studies of the solar photosphere and more realistic data-driven simulations of coronal magnetic field evolution.
\end{abstract}

\keywords{Sun: magnetic field, Sun: flares, Sun: sunspots}

\onecolumn
\thispagestyle{empty}

\tableofcontents

\newpage
\pagenumbering{arabic}

\section{{INTRODUCTION}}\label{intro}

The advent of high-cadence, large-scale vector magnetic field and
Doppler velocity measurements from instruments such as the Helioseismic
and Magnetic Imager (HMI, \cite{Schou2012}) on NASA's Solar Dynamics
Observatory (SDO) satellite \citep{Pesnell2012}, 
the Spectropolarimeter instrument (SP; \citealt{Lites2013}) on the Solar Optical Telescope \citep{Tsuneta2008} aboard the {\em
  Hinode} satellite \citep{Kosugi2007},
%
and improved capabilities of ground-based instruments, such as SOLIS
(e.g., \citealt{Keller2003}), make the estimation of electric fields
in the solar photosphere possible. The calculation of the electric
field from magnetic and Doppler data is critically important for
various quantitative studies of the solar atmosphere. First, if
we know both electric and magnetic field vectors in the photosphere,
we can estimate both the Poynting flux of magnetic energy and the flux
of relative magnetic helicity entering the corona. Second, as
demonstrated in a magneto-frictional model by \citet{Cheung2012}, the
ability to compute the electric field enables the driving of
time-dependent simulations of the coronal magnetic field from
photospheric magnetogram sequences. Combining electric field
estimation with a magneto-frictional model of the evolving solar
corona is the goal of the Coronal Global Evolutionary Model
(CGEM) project \citep{Fisher2015a}.\footnote{\url{http://cgem.stanford.edu}}



\cite{Kazachenko2014a} modified and extended the electric field
inversion methods introduced by \citet{Fisher2010, Fisher2012}, to
create a comprehensive technique for calculating photospheric electric
fields from vector magnetogram sequences. The new method, which we
dubbed the \texttt{PDFI} (an abbreviation for {\bf P}oloidal-Toroidal
Decomposition [PTD]-{\bf D}oppler-{\bf F}ourier Local Correlation
Tracking [FLCT]-{\bf I}deal) technique, incorporates Doppler
velocities from non-normal viewing angles (which are relevant to most
solar observations) and a faster and a more robust Poisson equation
solver, for obtaining the PTD solutions. After systematic, quantitative
tests of the accuracy and robustness of the \texttt{PDFI} technique,
using synthetic data from anelastic MHD (\texttt{ANMHD}) simulations
\citep{Abbett2004}, 
we found that the \texttt{PDFI} method has less than $1\%$ error
in the total Poynting flux and a $10\%$ error in the helicity flux
rate if we reconstruct it at the normal viewing angle ($\theta=0$) and
less than $25\%$ and $10\%$ errors respectively at large viewing
angles ($\theta=60^\circ$) \citep{Kazachenko2014a}.
 
In this paper, we take the next step forward, and apply the
\texttt{PDFI} technique to observations.  The flare-productive active region (AR) NOAA 11158 was observed by HMI nearly continuously for a
six-day period over 10-16 February 2011, starting from its initial emergence
near 14:00 UT on 10 February.
We use the sequence of magnetic and Doppler field measurements of NOAA 11158 to derive the temporal evolution of electric field, Poynting,
and helicity fluxes during these six days.
The evolution of this AR included two large bipoles emerging in close
proximity, with strong shearing motion between the central sunspots
\citep{Schrijver2011,Sun2012}. Over a six-day period, the AR hosted
an X2.2 flare (with the GOES soft X-ray flux peaking at 01:56 UT
on February 15) leading to a pronounced halo CME, three M-class
flares, and over twenty C-class flares.  Since this active region was
the first one for which the HMI vector magnetic field data were widely
distributed to the scientific community, its magnetic field has been
thoroughly studied:
the fast sunspot rotation from $20$ hours before to $1$ hour after the
X2.2 flare \citep{Jiang2012,Vemareddy2012a}, the flare related
enhancement in the horizontal magnetic field along the magnetic
polarity inversion line (PIL; \cite{Gosain2012, Wang2012, LiuC2012}),
abrupt changes in the vertical Lorentz force vectors
\citep{Petrie2012, Alvarado2012} and horizontal Lorentz forces
\citep{Petrie2013, Wang2014}, the evolution of relative and current
magnetic helicities \citep{Jing2012}, the injection of oppositely
signed helicity through the photosphere \citep{Dalmasse2013}, the subsurface three-dimensional magnetic
structure \citep{Chintzoglou2013}, the magnetic and velocity field
transients driven by the flare \citep{Maurya2012}. Numerous
approaches have been used to calculate the energy associated with X2.2
flare and the AR as a whole: the DAVE4VM method \citep{Liu2012,
  Tziotziou2013}, non-linear force free extrapolation \citep{Sun2012,
  Tziotziou2013}, the Minimum Current Corona Model \citep{Tarr2013},
and the coronal forward-fitting method \citep{Aschwanden2014,
  Malanushenko2014}. In this paper, we apply the \texttt{PDFI}
technique \citep{Kazachenko2014a} to derive electric fields at the
photosphere, and use those to estimate the photospheric energy and
helicities fluxes' evolution in NOAA 11158.

The paper is structured as follows. In Section~\ref{ptdtheory} we
briefly review the \texttt{PDFI} method itself and its recent
improvements.  In Section~\ref{data} we describe the observations of
the emerging, flaring NOAA 11158, and quantify how the uncertainties in
the HMI observations propagate into derived electric field and
Poynting fluxes. In Section~\ref{results}, we describe derived
electric fields, Poynting, and helicity fluxes in NOAA 11158, and in
Section~\ref{disc} we discuss results and draw conclusions. In
addition, in Appendix~\ref{merc} we show how our Mercator-reprojected
magnetic fields in Cartesian coordinates can be scaled to apply to
relatively small regions on the surface of the Sun to derive the
electric fields in a local Cartesian coordinate system.
\section{METHODOLOGY: \texttt{PDFI} TECHNIQUE, POYNTING \& HELICITY 
FLUXES}\label{ptdtheory}

The \texttt{PDFI} 
technique uses the evolution of the vector
magnetic field $\bvec$ and horizontal and Doppler velocities $\vvec$ to estimate
electric fields in the solar photosphere. It is described in detail in
\S2 of \citet{Kazachenko2014a}. The \texttt{PDFI} method combines the {\it
  inductive} contribution to the electric field ${\bf E}$ from
solution to Faraday's law using the Poloidal-Toroidal-Decomposition
(PTD) technique \citep{Fisher2010,Fisher2012}, with {\it
  non-inductive} contributions from $(-{\bf V} \times {\bf B})$. To
find the velocity ${\bf V}$, we use Doppler measurements and the Fourier
Local Correlation Tracking (\texttt{FLCT}) technique. The
\texttt{PDFI} technique is tested and its accuracy is
characterized in detail in \S4 of \citet{Kazachenko2014a}. In this
section we briefly describe the basics of the \texttt{PDFI} technique.

The fundamental idea of the \texttt{PTD} part of the \texttt{PDFI}
method is that the magnetic field, ${\bf B}$, defined on the
photospheric surface, has a solenoidal nature and thus can be
specified by two scalar functions $\mathcal{B}$ and $\mathcal{J}$:
\begin{equation}\label{eq1}
 {\bvec}=\nabla \times \nabla \times \mathcal{B}  {\bf \hat{z}} + 
\nabla \times \mathcal{J}  {\bf \hat{z}},
\end{equation}
where $\hat{z}$ points upward from the photosphere. Taking a partial
time derivative of Equation~(\ref{eq1}), and demanding that $\bvec$ obeys
Faraday's law, 
\begin{equation}\label{eq1.5}
 \frac{\partial {\bvec}}{\partial t} = -(\nabla \times c {\bf E}),
\end{equation}
we find a solution for the inductive part of the electric field ${\bf
  E^P}$ (where ``P'' stands for PTD), in terms of the partial time
derivatives of the poloidal and toroidal potentials,
$\dot{\mathcal{B}}$ and $ \dot{\mathcal{J}}$ respectively:
\begin{equation}\label{eq2}
c{\bf E^P}=- \nabla \times \dot{\mathcal{B}} {\bf \hat{z}} 
-\dot{\mathcal{J}} {\bf \hat{z}}.
\end{equation}
As described in \citet{Kazachenko2014a}, solving two-dimensional Poisson equations in the domain, where we
observe the vector magnetic fields, we determine $\dot{\mathcal{B}}$,
$\dot{\mathcal{J}}$ and $\frac{\partial \dot{\mathcal{B}}}{\partial
  z}$. Note  that the vector magnetic field data completely specify
the source terms of these Poisson equations (see \S2.1 of
\citet{Kazachenko2014a}).

The total electric field $\evec$ is a combination of the inductive and
non-inductive parts,
%
\begin{equation}\label{eq7}
c  {\evec}=- \nabla \times \dot{\mathcal{B}} {\bf \hat{z}} 
-\dot{\mathcal{J}} {\bf \hat{z}}-\nabla \psi 
\equiv \underbrace {c {\bf E^{P}}}_{\mbox {inductive}}-
\underbrace {\nabla \psi .}_{\mbox {non-inductive}}
\end{equation}
The non-inductive components to the scalar-potential part of the
solution include three separate contributions: (1) $-{\bf V} \times
{\bf B}$ from Doppler measurements (the ``D'' in \texttt{PDFI}), (2)
$-{\bf V} \times {\bf B}$ from Fourier Local Correlation Tracking
results (the ``F'' in \texttt{PDFI}), and (3) a scalar potential
contribution added to impose the constraint $\evec \cdot \bvec=0$,
consistent with the ideal MHD Ohm's law (the ``I'' in
\texttt{PDFI}). When adding the $-{\bf V} \times {\bf B}$
contributions above, any inductive contributions from these terms are
removed, since all inductive contributions are already included in the
${\bf E^{P}}$ solution.
Our approach for handling these non-inductive contributions is
described in detail in \citet{Kazachenko2014a}.

To calculate the flux of electromagnetic energy at the photosphere,
given by the Poynting flux vector
\begin{equation}
\label{eq_svec}
{\bf S}=\frac{c}{4\pi}({\bf E}\times {\bf B}),
\end{equation}
we use the observed magnetic field vector and the electric field
vector derived using the \texttt{PDFI} method. Since we are interested
in the amount of energy flowing into and out of the corona, we focus
most of our attention on the vertical component of Poynting flux,
\begin{equation}
\label{eq_sz}
S_z= \frac{c}{4\pi}\left(E_x B_y-E_y B_x\right).
\end{equation}
This depends upon the horizontal components of both the electric field
and the magnetic field. We further decompose $S_z$ into two
contributions, the flux of potential-field energy, and the flux of
free magnetic energy.  The basic idea is that the horizontal magnetic
field ${\bf B}_h$ can be divided into a potential-field contribution
${\bf B}_h^P$, and a contribution, ${\bf B}_h^f$, due to currents that
flow into the atmosphere from the photosphere \citep{Welsch2006}:
$S_z= \frac{c}{4\pi}{\bf E}_h \times \left({\bf B}_h^P+{\bf
  B}_h^f\right)$.  In this paper we use the Green's function to find
the potential field contribution, and subtract this contribution from
the measured horizontal fields to find ${\bf B}_h^f$. More discussion of the potential and free energy decomposition can be found in
\S3.1 of \cite{Kazachenko2014a}.

To calculate helicity flux rates, we use Equation~(62) from  \cite{Berger1984}:
\begin{equation}
\left(\frac{dH_R}{dt} \right) =
-2 \int \left({\bf A^P}\times {\bf E} \right) \cdot {\bf \hat{z}} \, da=
 \underbrace{-2 \int \left(A^P_{x}E_y-A^P_{y}E_x\right)da}_{\mbox{PDFI}},
\label{eq_hele}
\end{equation}
where ${\bf A^P}=\left(\frac{\partial \mathcal{B}^P}{\partial y},
-\frac{\partial \mathcal{B}^P}{\partial x},0\right) = \nabla \times B^P
\veczhat$ is the vector potential that generates the potential field
${\bf B^P}$ in volume $V$ above the photosphere, which matches the photospheric normal field $B_z$ at $z=0$.
(Note that a similar expression for helicity, Equation~(41) in
\citet{Kazachenko2014a}, contains a typographical error -- $E_z$ should be replaced by
$E_x$.) 
%
Here $B^P$ can be found by solving the Poisson equation A5 in
\citet{Fisher2010}. Adopting the naming convention from \S2.3.4 in
\cite{Kazachenko2014a}, the total helicity flux rate derived from the
\texttt{PDFI} electric field ${\bf E}$ will be referred to as the 
\texttt{PDFI} helicity flux rate or $\left(\frac{dH_R}{dt}
\right)_{PDFI}$ .

If we have an ideal electric
field $c{\bf E}=-{\bf V}\times{\bf B}$, then the helicity flux rate becomes \citep{Berger1984_rcc}: 
\begin{equation}
\left(\frac{dH_R}{dt} \right) =
\underbrace{- 2\int   [({\bf A^P}\cdot {\bf V_h})  B_z-({\bf A^P}\cdot {\bf B_h} ) V_z ]  da}_{\mbox{DFI}}.
\label{eq_helv}
\end{equation}
For observations near disk center, where the line-of-sight direction
approximates the vertical direction, in Equation~(\ref{eq_helv}), we can make the assumption that ${\bf V_h}$ can be determined with our FLCT flow estimates, and $V_z$ from our Doppler velocity measurements. Adopting the naming convention from \S2.3.4 in \cite{Kazachenko2014a}, the total helicity derived this way
would correspond to \texttt{DFI} electric field solution ({\bf D}oppler {\bf F}LCT {\bf
  I}deal).  

When comparing helicity fluxes calculated using the \texttt{PDFI} and
\texttt{DFI} techniques, it is important to remember that \texttt{PDFI} and
\texttt{DFI} methods are independent of each other, hence their results are not
necessarily consistent. 
For more details on \texttt{PDFI} and \texttt{DFI}
helicity fluxes and the quality of their reconstruction using
\texttt{ANMHD} simulations, see \S3.2 in \cite{Kazachenko2014a}.

\section{DATA REDUCTION: NOAA 11158} \label{data} 

We derive the evolution of the magnetic field, electric field, Poynting,
and helicity fluxes in NOAA 11158 using series of HMI vector
magnetograms and the \texttt{PDFI} method. A six-day, uninterrupted,
12-minute-cadence data set allowed us to study in detail both
long-term, gradual evolution, as well as rapid changes centered around the
region's X-class flare.  In this section we describe the data set and
the coordinate system re-projection we use.




NOAA 11158 was the source of an X2.2 flare on 2011/02/15, starting at
01:44 UT, peaking at 01:56 UT, and ending at 02:06 UT. A front-side
halo CME accompanied the flare \citep{Schrijver2011}.  Prior to the
X2.2 flare, the largest flare in this region was an M6.6 on 2011/02/13
at 17:28 UT, a little more than 30 hours before the X2.2 flare.


HMI
observed AR 11158 in great detail, routinely generating filtergrams in
six polarization states at six wavelengths on the Fe I 617.3 nm
spectral line. From these filtergrams, images for the Stokes
parameters, I, Q, U, and V were derived which, using the Very Fast
Inversion of the Stokes Algorithm (VFISV) code \citep{Borrero2011},
were inverted into the magnetic field vector components.  To resolve
the $180^\circ$ azimuthal field ambiguity the ``minimum energy''
method \citep{Metcalf1994, Leka2009} was used. In addition, we flipped
the azimuths of the transverse magnetic field vectors in all pixels
which exhibited single-frame fluctuations in azimuth of larger than
$120^{\circ}$ and for which such flipping reduced time variation in the azimuth \citep{Welsch2013}.

\subsection{Deriving Magnetic Vector Fields: ${\bf B}$}

To study preflare photospheric magnetic evolution, and to baseline
this evolution against postflare evolution, we obtained 153 hours of
12-minute-cadence 0.5$\arcsec$-pixel HMI vector-magnetogram data, from the 
beginning of the active region emergence, about four days before
the X2.2 flare, to two days after the flare: $t_{start}=$ February 10
2011 14:00 UT, with the active region centered at (S19, E50); and
$t_{end}=$ February 16 2011 23:48 UT, with the active region at (S21,
W37).

We rotated the active region 
to disk center and transformed it to a local Mercator re-projected Cartesian coordinate system \citep{Welsch2013}.  To do so, in the first step, we re-projected the observed magnetic vectors' components onto
radial/horizontal coordinate axes.
We then converted the Cartesian output grid's points into plane-of-sky
(POS) coordinates, and interpolated the radial and horizontal
components of the magnetic field, $B_r$ and $\bvec_h$, onto the remapped
output grid points. Finally, to account for small, whole-frame shifts of the AR's
structure between successive measurements, we co-aligned the data to sub-pixel scale.
Note that since FLCT \citep{Fisher2008}, and indeed any method of estimating the
optical flow (e.g., \cite{Schuck2006}), depends upon image structure
(e.g., gradients), conformal mappings are preferred since they are
shape-preserving for infinitesimally small objects.  Accordingly, we use Mercator re-projection
\citep{Welsch2009}, with equally spaced grid points as an input for
\texttt{FLCT}. After re-projection, to preserve physical quantities of
magnetic fields and velocities, we corrected the flux densities for the
distortion of pixel scale introduced by re-projection; the details of
the applied correction-factors are given in Appendix~\ref{merc}.

For the minimum magnetic field to consider in the PTD, we chose a
threshold of $|{\bf B}|=250$ Gauss, consistent with the upper limit of
the uncertainty in the horizontal and vertical components of the
magnetic field \citep{Hoeksema2014}. To avoid spurious signals in electric
fields, we apply a mask, where we set any pixel's magnetic field
components to zero if in any of three consecutive frames it has $|{\bf
  B}|<250$ Gauss.  To increase the accuracy of the calculated electric
fields we also added a boundary area of 55-pixels width/height padded
with zeroes around the periphery of the magnetogram (see \S2.1 of
\citet{Kazachenko2014a}).  The final data cube, after re-projection
and boundary padding, consists of 770 time steps ($dt=720$ sec) and has
a field of view of $665\times645$ pixels with a pixel size of $360.16$
km, which at disk center is equivalent to the original 0.5'' size of
HMI's pixels.  Further detail on the data cube preparation and
calibration, for a shorter time period, can be found in
\cite{Welsch2013}.

\begin{figure*}[ht!]
  \centering 
  \resizebox{1.0\hsize}{!}{\includegraphics[angle=0]{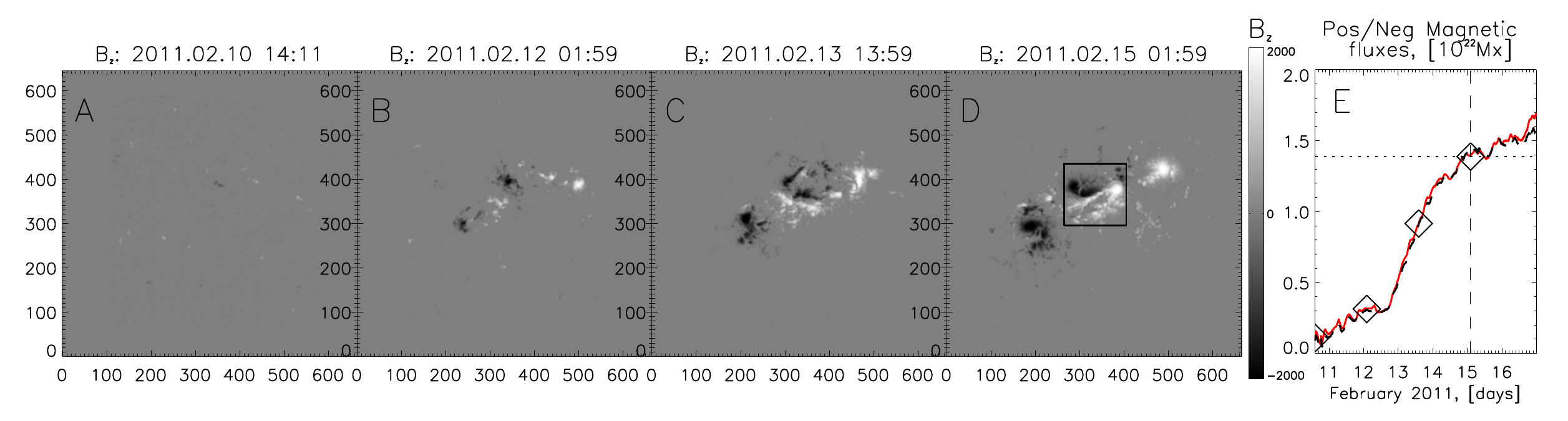}}
  \caption{{\it Panels A-D:} HMI vertical magnetic field ($B_z$) maps
    at 4 different times of NOAA 11158 evolution. {\it Panel E}:
    evolution of the positive and negative vertical magnetic fluxes of
    the 6-day interval, with the diamonds indicating the times of the
    four images (A-D) on the left. An X-class flare occurred at the
    time corresponding to the vertical dashed line. The black box in
    {\it Panel D} indicates the field of view of
    Figures~\ref{fig_bh}--\ref{fig_eszanal}, however all energy and helicity estimates 
    described further are for the entire field of view.}
  \label{fig_mag}
\end{figure*}

Figure~\ref{fig_mag} shows the final vertical magnetic field in a
subregion of the full-disk data array after re-projection in the
beginning (Panel $A$), middle ($B,C$) and the postflare ($D$) times
of the magnetogram sequence.  Note that the positive and negative
vertical magnetic fluxes, $B_z(t)$, shown on the right panel, nearly
balance each other; the signed magnetic fluxes grow from essentially zero
to roughly $1.4\times10^{22}$ Mx ( $-1.4\times10^{22}$ Mx) at the time of the flare (vertical
dashed line). The flux emerges in two phases -- an initial, gradual 
phase is followed by a much more rapid phase, a pattern seen in
the emergence of many active regions \citep{Fu2015}.

\subsection{Deriving Velocity Vector Fields: ${\bf V}$}

To derive the three-component velocity vector of the magnetized
plasma, we used the following two methods: the method of 
\cite{Welsch2013} for calibrated line-of-sight (LOS) Doppler velocity
component $V_{LOS}$, and \texttt{FLCT} for the horizontal velocities $V_h$.
To derive and calibrate the Doppler velocity for instrumental effects and convective blueshift, we used three successive
vector magnetograms and one Dopplergram coincident with the central
magnetogram. Following the idea of \cite{Welsch2013}, that the Doppler
shifts measured along polarity inversion lines (PILs) of the LOS
magnetic field determine one component of the velocity perpendicular
to the magnetic field,  we calibrated the quantity $V_{LOS}$ by subtracting the median Doppler velocity among all pixels on LOS PILs.
To find the horizontal velocity $V_h$, we determined local displacements of
magnetic flux between two successive images in the neighborhood of
each pixel, employing the following three steps. First, we masked the
initial and final images with a Gaussian windowing function with an
$e$-folding width of $\sigma_{FLCT}=5$ pixels; second, we
cross-correlated the two masked images; finally, we found the peak of
the cross correlation function. The vector displacement of this peak from zero
is the inferred spatial displacement of the pattern in the
neighborhood of the windowing function's center.

To calculate electric fields, and hence Poynting and helicity fluxes (Equations~(\ref{eq_sz}) and (\ref{eq_hele})), we use Doppler and \texttt{FLCT} velocities as an input into the
\texttt{PDFI} inversion. For comparison, apart from the \texttt{PDFI},
we also use Doppler and \texttt{FLCT} velocities on their own as an
independent estimate for the helicity flux rate (see
Equation~(\ref{eq_helv})).


To summarize, as a result of the data reduction we obtained a six-day
data cube consisting of 768 frames (two frames less than the original dataset to obtain \texttt{FLCT} velocities), each of which contains data for three components of the velocity field and three components of the
magnetic field with a field of view of $665\times645$ pixels, a pixel
size of $360.16$ km, and a time step of $dt=720$ sec.


\subsection{How do Errors in Magnetic Field Measurements 
Affect our Electric Fields Estimates?}
\label{errors}



When testing the \texttt{PDFI} electric field inversion technique
using MHD simulations \citep{Kazachenko2014a}, we have the advantage
of knowing that the errors in the input data are zero, but such is not
the case with the HMI data \citep{Liu2012a}.  Hence to estimate the
uncertainties in the derived electric field and Poynting fluxes, we
have to account for uncertainties in the HMI input data,
which arise primarily from estimation and inversion of the Stokes
profiles \citep{Liu2012a,Hoeksema2014}. 



\begin{figure*}[htb!]
  \centering 
  \resizebox{0.5\hsize}{!}{\includegraphics[angle=0]{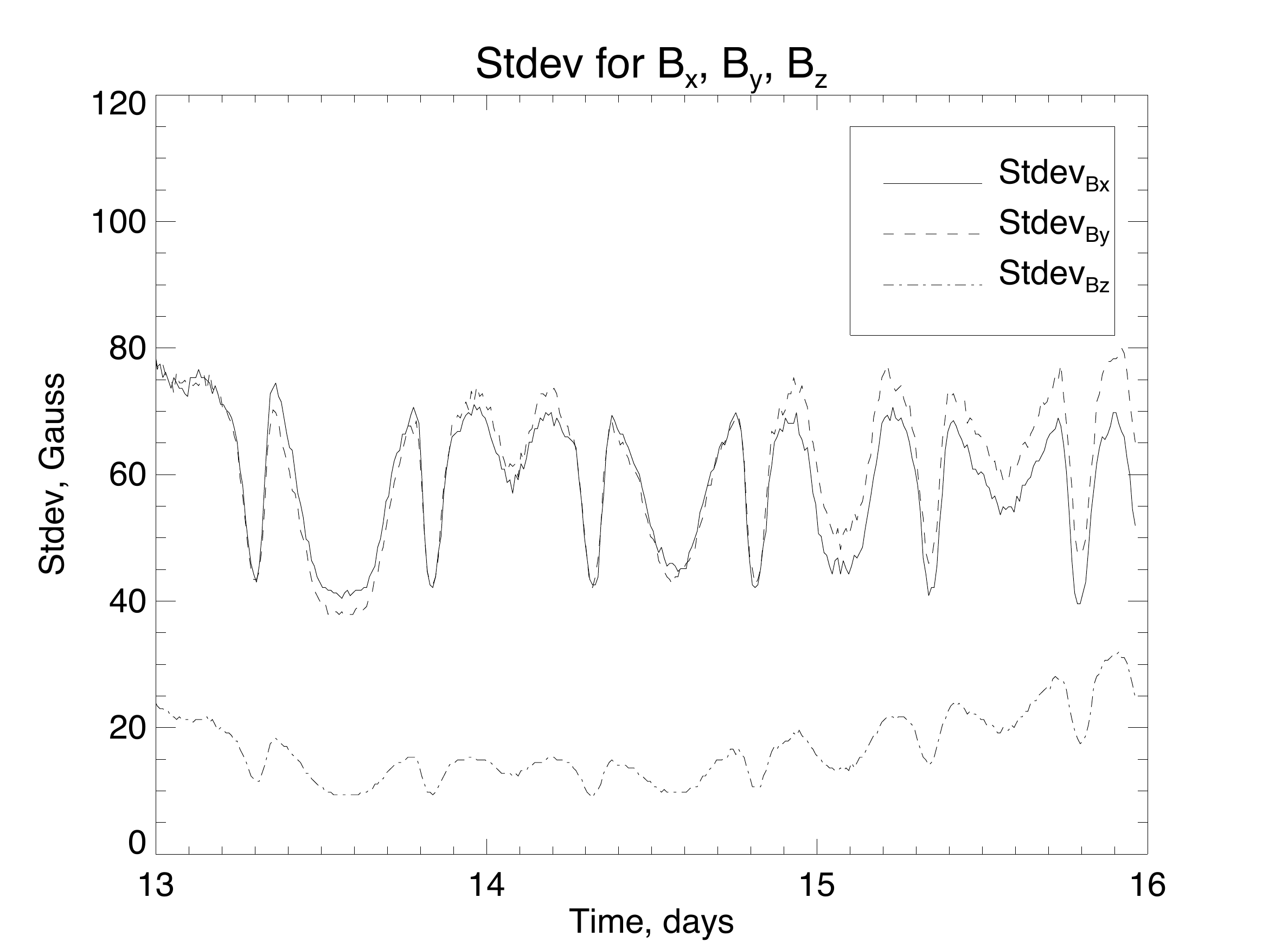}}
  \caption{Estimated noise levels in $B_x$, $B_y$ and $B_z$, as
    functions of time in February 2011. Note the periodic variation,
    believed to be a function of orbital phase of the SDO satellite. }
  \label{noisetest} 
\end{figure*}

Fitting the core of magnetic-field-values distribution in the weak-field regions with a Gaussian and assuming that its width indicates the noise level in the measurements, we estimate the errors in $B_x$, $B_y$ and $B_z$ in AR 11158 dataset during six days of AR evolution (see Figure ~\ref{noisetest}). The fluctuation of the error in $B_x$, $B_y$ and
$B_z$ varies within 100 Gauss for the horizontal magnetic field and
within 30 Gauss for vertical magnetic field due largely to effects arising from SDO's 24-hour orbital period \citep{Hoeksema2014}. We use those values,
i.e., $[100,100,30]$ Gauss, respectively, as noise thresholds for
magnetic field values. 


 \begin{figure*}[htb!]
  \centering 
  \resizebox{0.9\hsize}{!}{\includegraphics[angle=0]{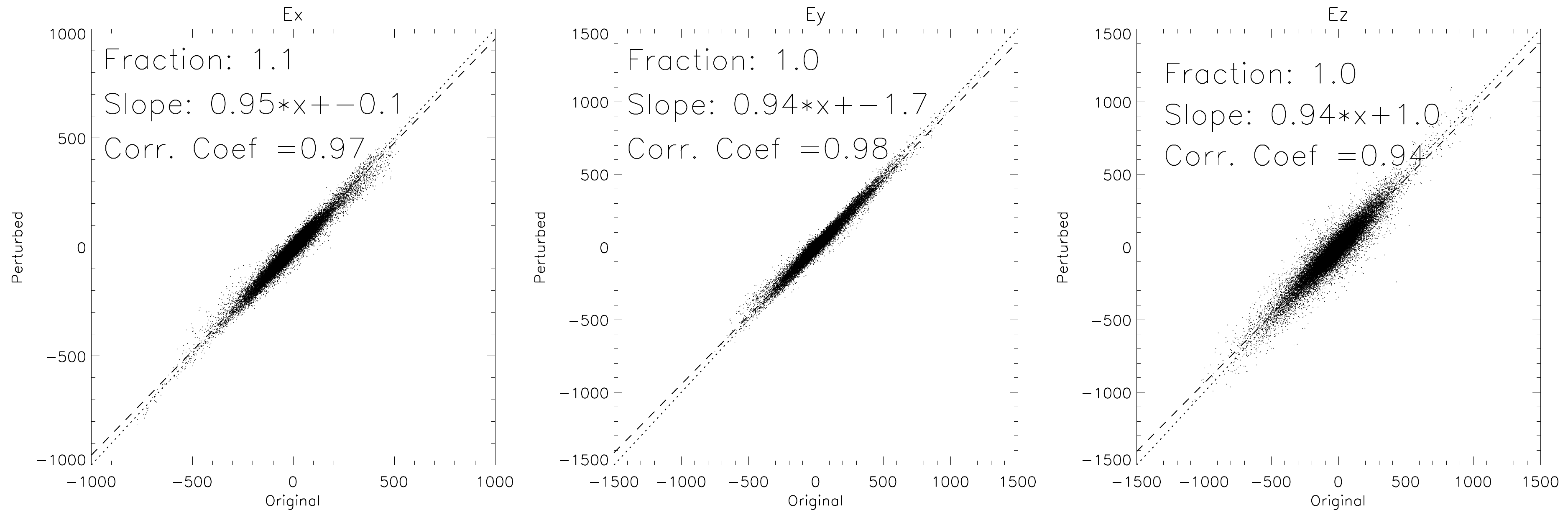}}
  \caption{Pixel-by-pixel scatter plots showing the ensemble of
    perturbed electric field solutions as a function of the
    unperturbed electric field solution, for all three components of
    {\bf E}, for one particular realization of pseudo-random noise.}
  \label{eperturb} 
\end{figure*}



We use Monte Carlo simulations to estimate the errors in the
\texttt{PDFI} electric fields caused by the uncertainty in the HMI
data. The sensitivity of \texttt{PDFI} electric field solutions to
vector magnetogram noise is exacerbated by the fact that source terms of the Poisson equations that are
%
solved as part of 
the electric field inversions involve temporal and/or spatial
derivatives, which greatly amplify the noise.  This is, however,
ameliorated by the fact that solutions to Poisson's equation tend to
smooth out the effect of noisy source terms. Since the entire
inversion procedure is quite complex, our approach is to start from a
given set of input magnetic field data, and then 
%
add pseudo-random, Gaussian perturbations to the data, consistent with
the noise thresholds given above ($[100,100,30]$ Gauss). 
By comparing the ensemble of electric field
solutions that are perturbed about our initial solution, we can
characterize the resulting errors of the electric field inversion.
Figure~\ref{eperturb}
shows three scatter plots between the original and the perturbed 
the electric fields for one particular realization of pseudo-random noise, one plot for each component.


The resulting uncertainties in $E_x$ and $E_y$ are smaller than those
for $E_z$, reflecting the fact that the inductive horizontal electric
field from PTD, ${\bf E}_h^P$, depends on $B_z$, which has much
smaller intrinsic errors than $B_x$ and $B_y$, while the inductive
part of the vertical electric field, $E_z^P,$ depends on the
more-noisy $B_x$ and $B_y$. Thus our study not only shows that the
electric field inversion errors are not too large, but the
distribution of these errors about the diagonal line provides
quantitative estimates for the random errors, and deviations from unity
slope provide some information about more systematic errors.  To scale
the deviations relative to the unperturbed electric fields, we fit the
difference between the electric field from the perturbed magnetic
field and the electric field from the unperturbed magnetic field with
a Gaussian and then normalize it by dividing it by the standard
deviation of the unperturbed data. 
We find the standard deviation $\sigma=[0.14,0.13,0.18,0.14]$ for $E_x$, $E_y$, $E_z$ and $S_z$, respectively.  
%

To summarize, the levels of the errors in
the HMI data lead to errors in the electric field up to $14\%$ and
$13\%$ in $E_x$ and $E_y$ and $18\%$ in $E_z$.  The error in the
vertical Poynting flux, $S_z$, is $14\%$.  If we add these
uncertainties from the data (in quadrature) to the errors in the total
Poynting flux that we found when testing the \texttt{PDFI} method
(where the discrepancy, $\delta S_z$, ranged from $1\%$ to $25\%$ for
viewing angles ranging from $0$ to $60^\circ$, see Figure~10 of
\citet{Kazachenko2014a}), then we end up with the error ranging from
$14\%$ to $29\%$ in the vertical Poynting flux. 
\section{RESULTS}\label{results}

To describe the photospheric electric fields ${\bf E}$ and energy
fluxes ${\bf S}$ in evolving NOAA 11158, we use two approaches. In
Section~\ref{spatial} we show the spatial distribution of ${\bf B}$,
${\bf V}$, ${\bf E}$ and ${\bf S}$ at two times, before and after the
X2.2 flare.  In Section~\ref{temporal} we analyze their temporal
spatially integrated evolution over the six days of observations. These approaches allow
us to capture both the spatial structure at quiet times before the
flare with the changes during the flare, and the overall long-term
behavior of the active region.


\subsection{Properties Of the Active Region NOAA 11158 Before and After the X2.2 Flare}\label{spatial}

As an example of the spatial distribution of electric fields and
Poynting fluxes in NOAA 11158, we selected two instances before
$T_{pre}=01:36$ UT and after $T_{post}=02:12$ UT the X2.2 flare.
%
%
 These instances are of particular interest since the photospheric vector
magnetic field changed abruptly during the flare \citep{Petrie2012,Wang2014}, and various flare signatures, such as flare ribbons, have been observed during this time frame. 
Here we investigate the changes in velocities, electric fields, and
Poynting fluxes associated with these changes.
We note that the $12$-minute-cadence data shown here are derived from a tapered temporal average that is performed every $720$ seconds using observations collected over $22.5$ minutes ($1350$ s) to reduce noise and minimize the effects of solar oscillations \citep{Hoeksema2014}. 

\subsubsection{Magnetic Field: ${\bf B_{preflare}}$ and ${\bf B_{postflare}}$ } 

\begin{figure*}[htb!]
  \centering   \resizebox{1.05\hsize}{!}{\includegraphics[angle=0]{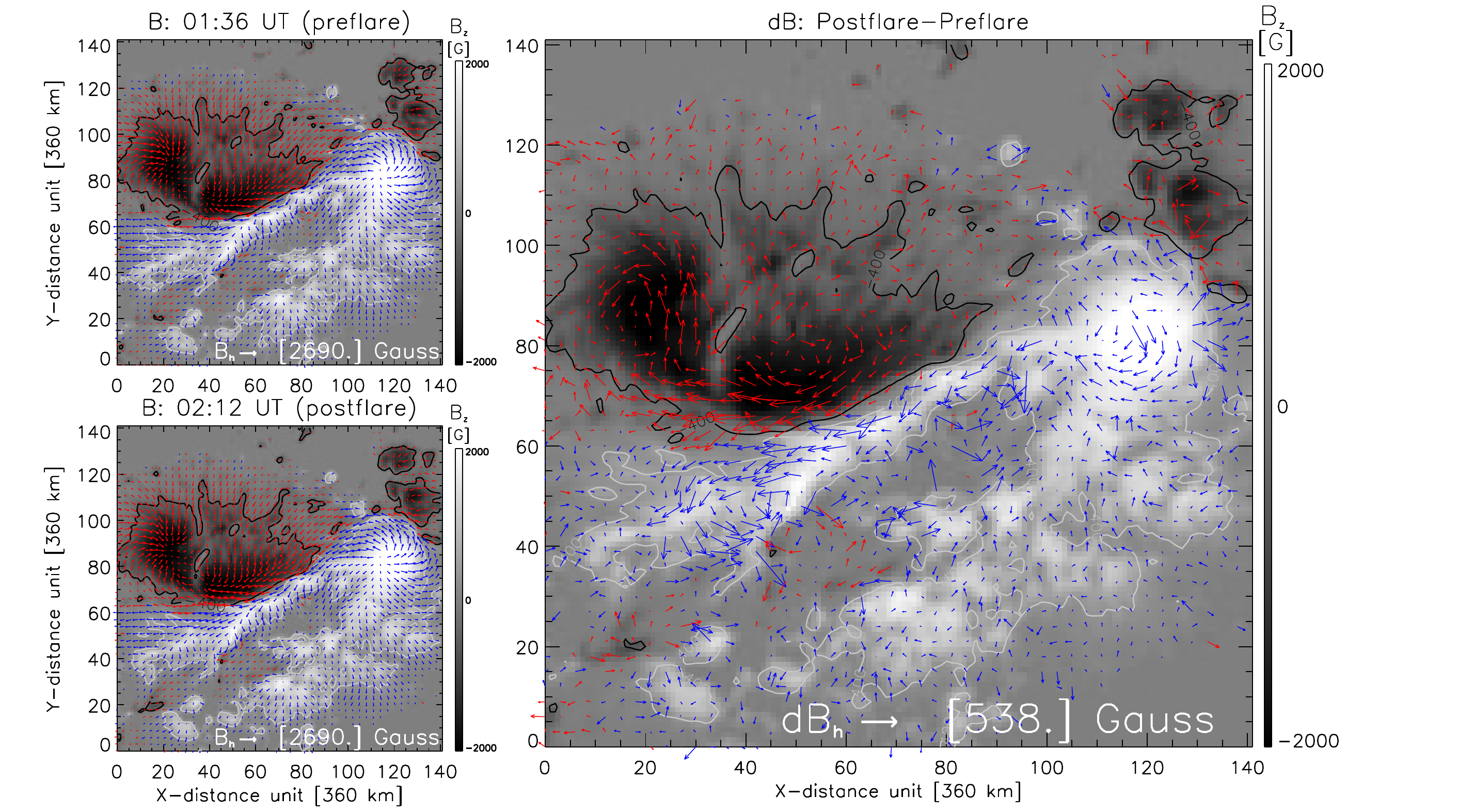}}
  \caption{Horizontal (arrows) and vertical (grayscale background)
    components of the magnetic field in NOAA 11158 at preflare ({\it
      top left}) and postflare times ({\it bottom left}), and the difference
    horizontal field between the two, postflare minus preflare ({\it right}
    panel). The background white and black colors represent positive
    and negative vertical magnetic fields, respectively. The blue and
    red colors correspond to horizontal field in areas of positive and
    negative values of the background vertical magnetic field
    $B_z$. The white and black contours outline the positive and
    negative vertical magnetic fields at $B_z=\pm400$ Gauss.  The
    arrows in the right bottom corners show scales for horizontal
    magnetic field vectors, ${\bf B}_h$ (left panels) or its
    change, ${\bf dB}_h$ (right panel).}
  \label{fig_bh}
\end{figure*}

Figure~\ref{fig_bh} shows the vector magnetic fields ${\bf B}$
centered at NOAA 11158 before (preflare, $T_{pre}=01:36$ UT) and the first time step after the onset of the flare (postflare, $T_{post}=02:12$ UT) and also the
difference image between the two (right panel). As seen from the difference image on the right, the horizontal magnetic field close to PIL increased by over $300$ G during the flare (see arrows), while the vertical magnetic
field remained nearly constant. This horizontal-field increase is
consistent with the magnetic field contraction scenario of
\cite{Hudson2008} and is described in detail by \citet{Petrie2012,
  Wang2014} and \citet{Sun2012}. On the difference image we also notice the two circular patterns,
directed counter-clockwise in negative (pixel coordinates [20,90]) and clockwise in positive (pixel coordinates [120,80])
polarities, meaning that the field connecting positive and negative polarities becomes more left-handed. Since both preflare and postflare magnetic fields have a right-handed twist  (i.e. horizontal magnetic fields have clockwise orientation in the negative polarity and counter-clockwise orientation in the positive polarity), the observed change, $dB$, decreases the twist present in the preflare magnetogram -- direct evidence of an abrupt magnetic twist decrease in both sunspots during the flare, which contradicts arguments \citep{Melrose1995} that the vertical current density through the photosphere should not change on the flare time scale. The sudden change in twist might arise from removal of magnetic
helicity from the active-region's magnetic field by a coronal mass
ejection \citep{Longcope2000c, Petrie2012}.

\subsubsection{Velocity Field: ${\bf V_{preflare}}$ and ${\bf V_{postflare}}$ } 

 \begin{figure*}[htb!]  
\centering
\resizebox{1.1\hsize}{!}{\includegraphics[angle=0]{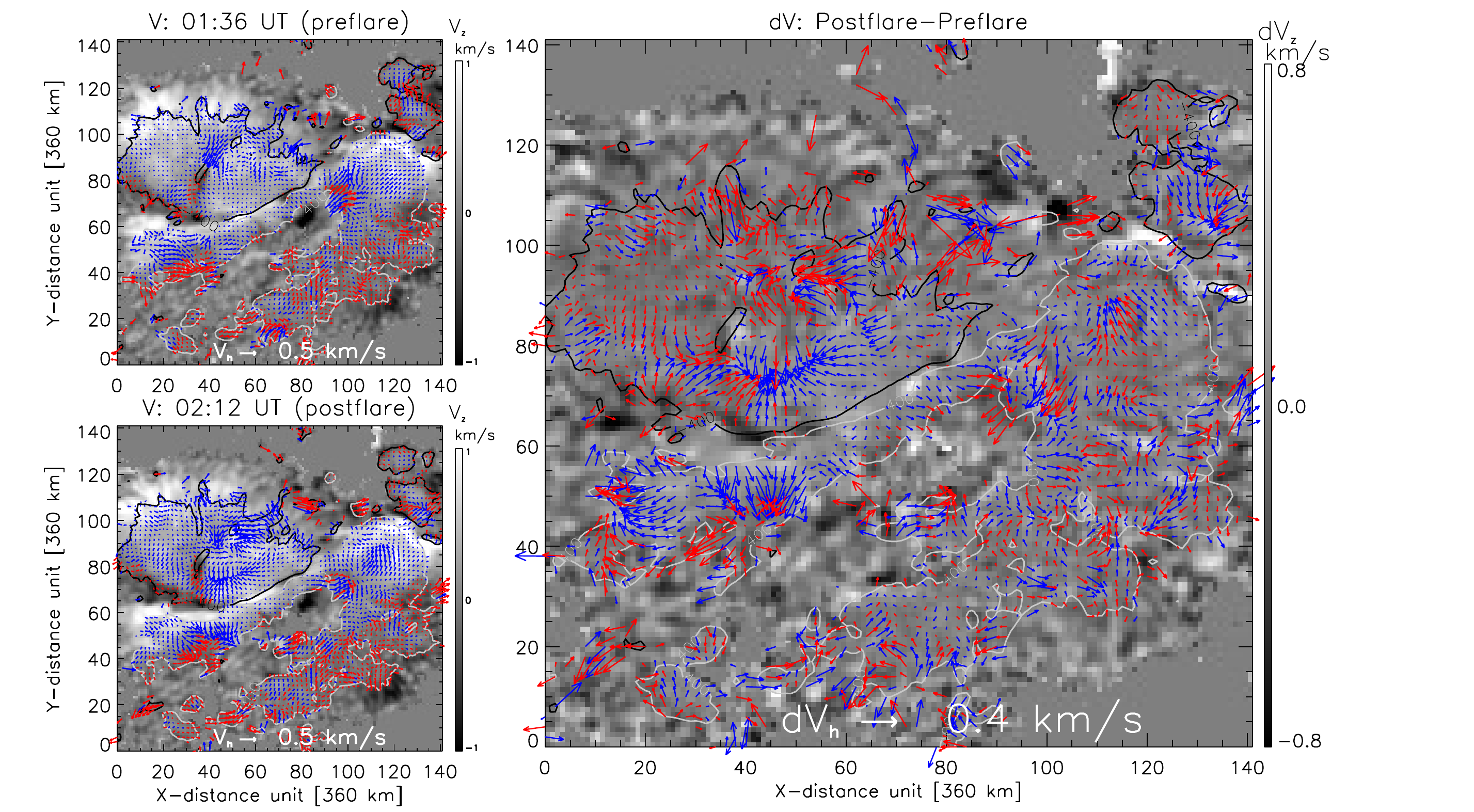}} 
\caption{Horizontal (arrows) and vertical (background) velocity field
  in NOAA 11158 at preflare ({\it top left}) and postflare times ({\it
    bottom left}), and the difference image between the two ({\it
    right} panel). The background white and black colors represent positive
  and negative Doppler velocities, respectively, where positive is
  toward the viewer (opposite to the usual astrophysical convention). The
  blue and red colors correspond to ${\bf V}_h$ originating in areas
  of positive and negative values of background vertical velocity. The
  white and black contours outline the positive and negative vertical
  magnetic fields at $B_z=\pm400$ Gauss. The arrows in the right
  bottom corners show scales for horizontal velocities.}
\label{fig_vel}\end{figure*}

For completeness, in Figure \ref{fig_vel} we show the three components
of the velocity field at the times before and after the flare as well
as the difference image between the two. The Doppler velocity has a
strong ($1$ km s$^{-1}$) upflow close to PIL and in the sunspot's
umbra. During the flare the Doppler speed does not exhibit any
prominent changes (see the background of the right panel), although
such changes have been reported in other flares \citep{Deng2006}.
However the horizontal velocity field does change: there is an
apparent $\sim 0.2$ km/s drift of magnetic field away from the PIL
during the flare. More details on the horizontal velocity pattern during the flare can be
found in \citet{Wang2014}, where $45$-second cadence HMI intensity
maps were used.


\subsubsection{Electric Field: ${\bf E_{preflare}}$ and ${\bf E_{postflare}}$ } 

\begin{figure*}[htb!]
  \centering   
  \resizebox{1.1\hsize}{!}{\includegraphics[angle=0]{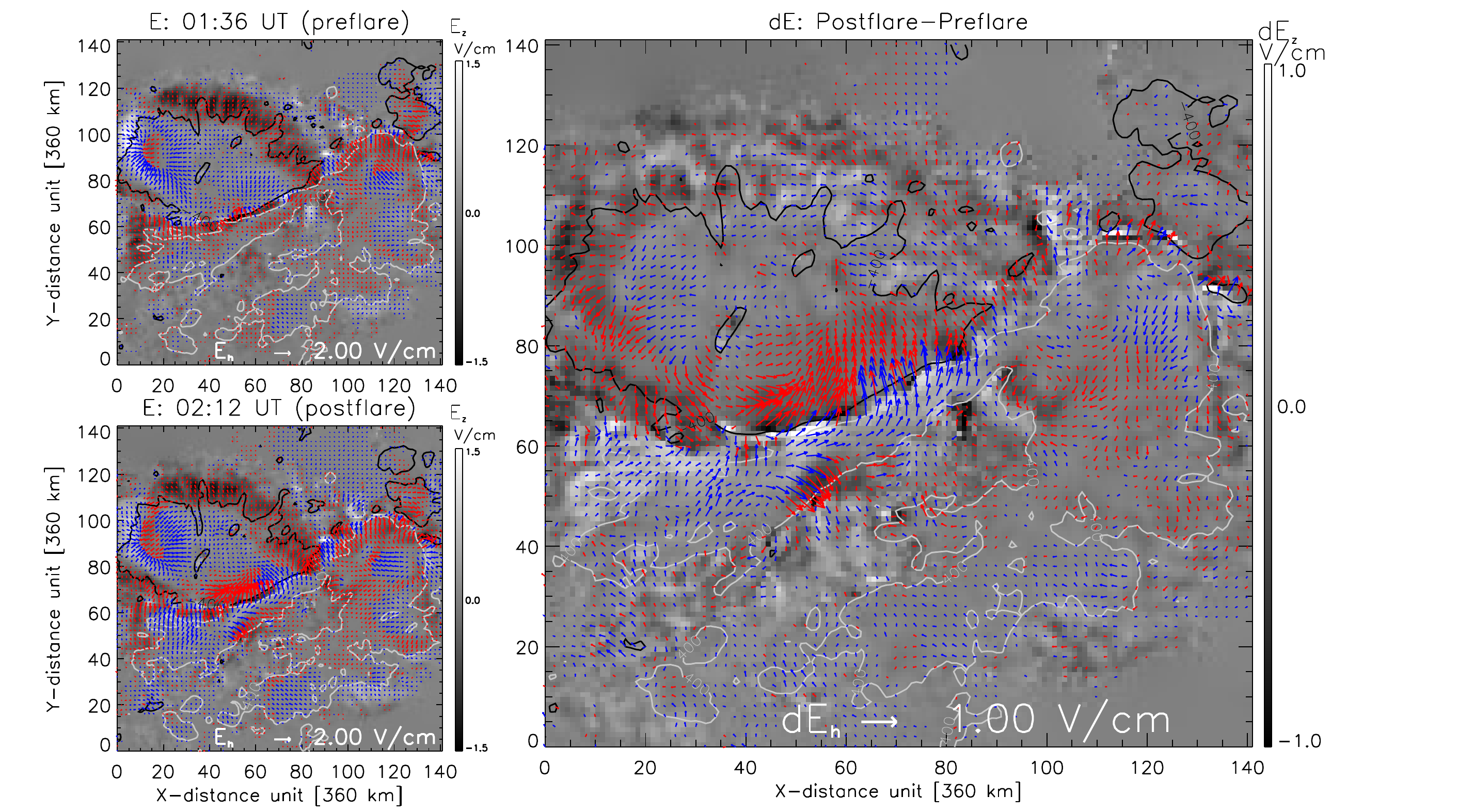}}
  \caption{``Electrogram'': Horizontal (arrows) and vertical
    (background) electric field components in NOAA 11158 at preflare
    and postflare times, and the difference image between the
    two. The blue and red colors correspond to ${\bf E}_h$ in areas of
    positive and negative background $E_z$. The white and black
    contours outline the positive and negative vertical magnetic
    fields at $B_z=\pm400$ Gauss.  The arrows in the right bottom
    corners show scales of horizontal vectors.}
  \label{fig_eh_bz}
\end{figure*}

Finally, in Figure~\ref{fig_eh_bz} we show the vector electric field
maps (``vector electrograms'') calculated using the \texttt{PDFI}
method before and after the flare, and the difference between the two. 
The horizontal electric fields ${\bf E}_h$ range from 0 to $1.2$ V/cm, with the strongest values concentrated close to the PIL. The strong ${\bf E}_h$ at the PIL is a consequence of the steady velocity upflows and large horizontal magnetic fields along the PIL that lead to 
a non-inductive, horizontal electric field, ${\bf E}_h^D$ --- the
``D'' in \texttt{PDFI} --- oriented in a perpendicular direction to the PIL.  Given the flare-driven
increase in the horizontal magnetic field at and near the flaring PIL,
this ${\bf E}_h^D$ and hence ${\bf E}_h$ also increase, by up to $0.5$ V/cm close to PIL, as shown in
Figure~\ref{fig_eh_bz}. As for the vertical component, $E_z$ varies within $[-2.2,1.8]$ V/cm. The black island at the PIL before and after the flare reflects a presence
of the strong vertical electric field of $E_z\approx-1.5$ V/cm directed into the Sun that remains nearly constant during the flare. What are
the physical consequences of the presence of a strong vertical
electric field? The spatial distribution of the vertical component of
the inductive electric field from PTD, $E_z^P$, is related to the time
derivative of vertical current density via $\partial_t J_z= c (\nabla^2 c E_z^P)/(4
  \pi)$. Hence, the presence of a nonzero $E_z$ is related to changes in the
vertical current, which is mostly concentrated close to the main PIL
\citep{Petrie2013, Janvier2014}.  
%
%

\subsubsection{Poynting Flux Vector: ${\bf S_{preflare}}$ and ${\bf S_{postflare}}$ } 
\label{szxy}

\begin{figure*}[htb!]  \centering  
\resizebox{1.1\hsize}{!}{\includegraphics[angle=0]{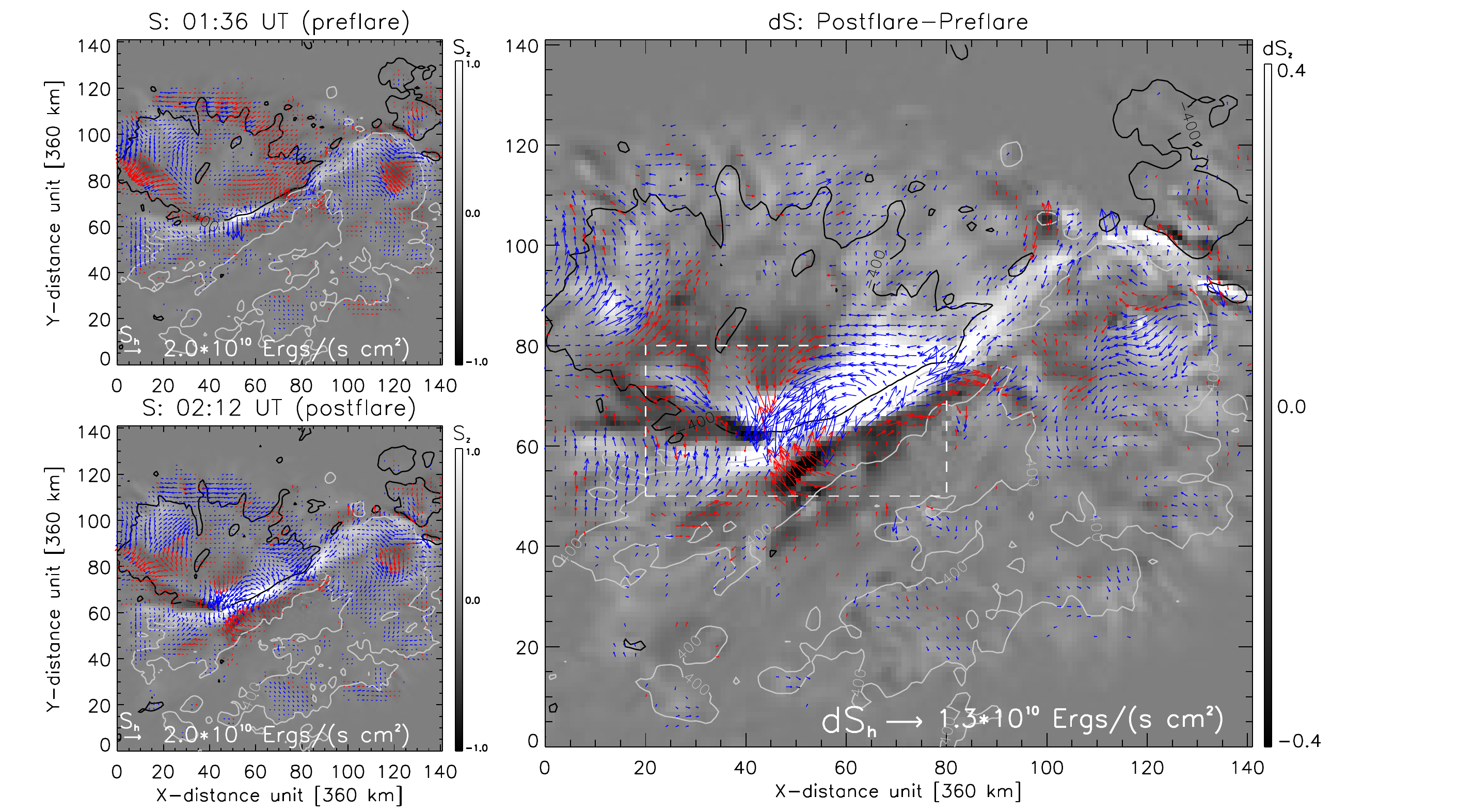}} 
\caption{Horizontal (arrows) and vertical (background) Poynting vector
  field components in NOAA 11158 at preflare and postflare times,
  and the difference image between the two.  The blue and red colors
  correspond to ${\bf S}_h$ in positive and negative areas of the
  background $S_z$. The white and black contours outline the positive
  and negative vertical magnetic fields at $B_z=\pm400$ Gauss.  The arrows in the
  right bottom corners show scales of horizontal vectors. The range of
  the background $S_z$ is $[-1,1]\times10^{10}$ ergs cm$^{-2}$ s$^{-1}$
  in the left panels and $[-0.4,0.4]\times10^{10}$ erg cm$^{-2}$
  s$^{-1}$ in the right panel. The {\it white dashed box} in the center of the right panel indicates the field of view where energy injection rates are calculated (see Section~\ref{szxy}).}  
\label{fig_sh}
\end{figure*}

Figure~\ref{fig_sh} shows snapshots of the three components of the
Poynting flux vector, ${\bf S}=\frac{c}{4\pi} {\bf E} \times {\bf B}$,
before and after the flare and also the difference image between the
two. The vertical Poynting varies within $[-0.6,2.3]\times10^{10}$~ergs~$\cdot~cm^{-2}~\cdot s^{-1}$. The
maximum value of this range is around three times smaller than the steady-state
photospheric energy flux estimated from the Stefan-Boltzmann law, given the temperature $T_{photosphere}=5500K$,  $f\sim
6\times10^{10}$~erg~$\cdot~$cm$^{-2}\cdot~ $s$^{-1}$. 
By analyzing the distribution of the vertical Poynting flux, we find that more than half of the spatially integrated signed Poynting flux both at preflare and postflare instants is injected in the range of $S_z=[10^9,10^{10}]$ ergs $\cdot cm^{-2}\cdot s^{-1}$, while the rest of the energy is injected in the range of $S_z=[10^8,10^9]$ ergs $\cdot cm^{-2}\cdot s^{-1}$ with the positive flux.  We find the strongest vertical Poynting fluxes  close to the main PIL: during the flare the mean values of positive and negative $S_z$ contributions within a white dashed box, shown on the right panel of Figure~\ref{fig_sh}, change from $(1.8$ to $4.5)\times10^9$ ergs $\cdot$ cm$^{-2} \cdot $s$^{-1}$ and from $(-0.6$ to $-1.4)$ $\times10^9$ ergs $\cdot$ cm$^{-2} \cdot $s$^{-1}$ respectively, or spatially integrating over the same box in terms of energy injection rate, from
$1.3\times10^{27}$ ergs s$^{-1}$ to $4.9\times10^{27}$ ergs s$^{-1}$ (see white
enhancement on the right panel). Summed over the 22 minutes of the
GOES flare time for this event, the energy crossing the photosphere during the flare within the box, would amount to
about $4.1\times10^{30}$ erg, that is very small compared to typical energies released in large flares such as this ($10^{32}$ erg
or more).  We also note that the absolute value of the spatially integrated positive Poynting flux (over NOAA 11158 ) is more that two times larger than the absolute value of the spatially integrated negative Poynting flux -- more magnetic energy is transported through the photosphere upward into the corona than downward.


Similar to the evolution in $E_h$, the magnitude of the horizontal
component of the Poynting flux, $S_h$, increases near the PIL by up to $8\times10^{9}$ ergs $\cdot$
cm$^{-2} \cdot $s$^{-1}$.  As noted above, the combination of steady
upflows found in the Dopplergrams at and near the PIL with the
systematic increase in the transverse magnetic fields along the PIL
leads to an increase in the non-inductive, horizontal electric field,
${\bf E}_h^D$, which in turn produces an increase in the
Poynting flux there.


\begin{figure*}[htb!]  \centering  
\resizebox{0.95\hsize}{!}{\includegraphics[angle=0]{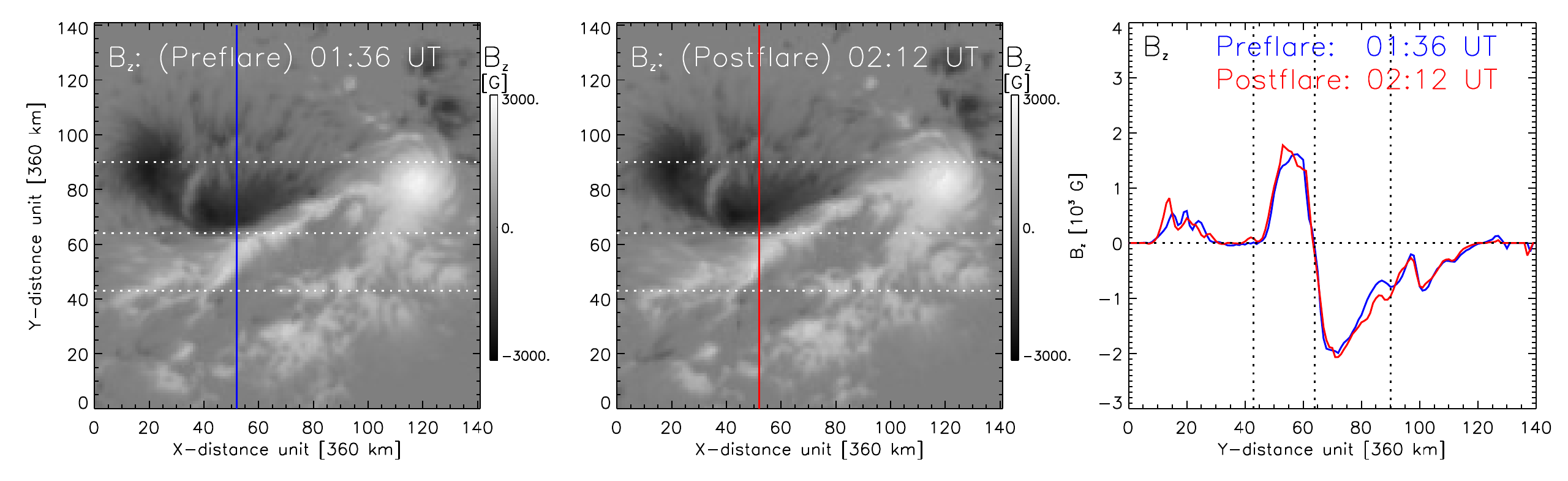}} 
\resizebox{0.95\hsize}{!}{\includegraphics[angle=0]{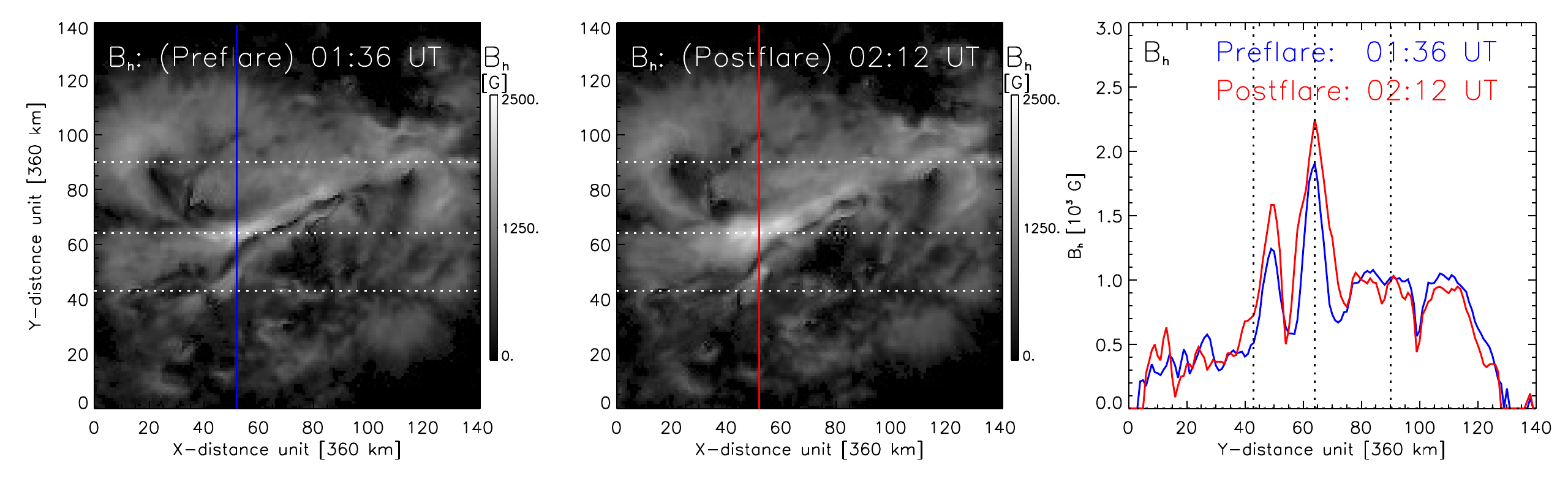}} 
\resizebox{0.95\hsize}{!}{\includegraphics[angle=0]{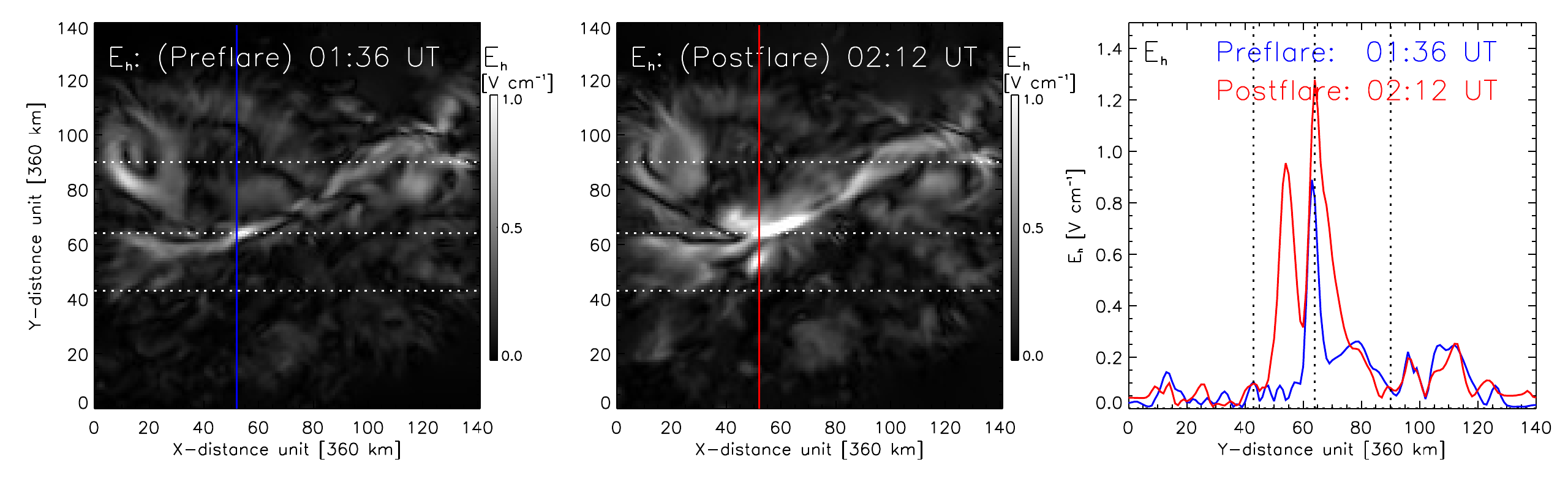}} 
\resizebox{0.95\hsize}{!}{\includegraphics[angle=0]{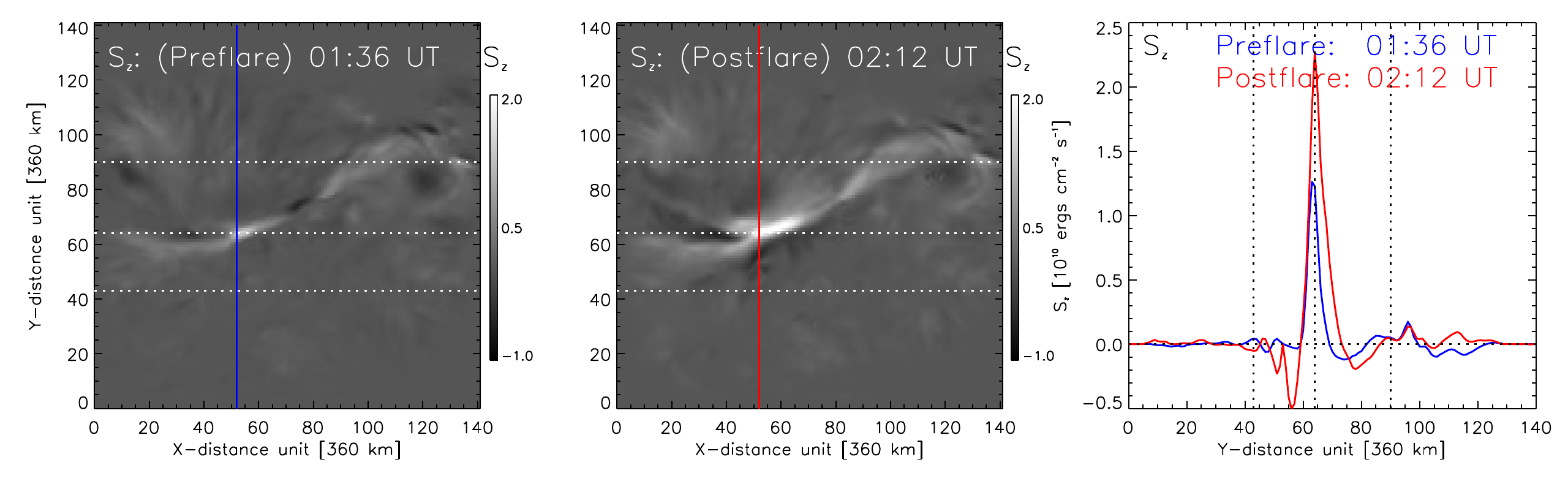}} 
\caption{{\it Left and Middle columns:} Vertical and horizontal
  magnetic fields, horizontal electric field, and the vertical
  Poynting flux before and after the flare. {\it Right column:}
  Vertical scans for $B_z$, $B_h$, $E_h$ and $S_z$ before (blue) and
  after (red) the flare at the locations shown with blue and red lines
  in the two left columns. Horizontal dotted lines on left and middle panels correspond to vertical dotted lines on the right panel.}  \label{fig_eszanal}\end{figure*}

Summarizing some of the changes in the photosphere during the flare,
in Figure~\ref{fig_eszanal}, we show the vertical and horizontal
magnetic fields, horizontal electric field and the vertical Poynting
flux before and after the flare and also the scans along a fixed $x$ value (or column in the 2D image) through
the center of the active region across the PIL at those two moments (right column).
From the right panel we find that while the vertical magnetic field does not exhibit
any significant changes (see $B_z$-panel, first row), the horizontal magnetic field increases at the PIL by
over $300$ Gauss (see $B_h$-panel, second row). This leads to an increase in the horizontal
Doppler electric field by up to $0.5$ V/cm close to PIL and by almost $1$ V/cm in some locations away from the PIL (see blue and red plots for $E_h$ close to vertical dotted lines, third row) and an increase in the 
vertical Poynting flux close to PIL by up to $10^{10}$
ergs$\cdot$cm$^{-2}$s$^{-1}$, from $[1.25$ to $2.25]$ $\times10^{10}$ ergs$\cdot$cm$^{-2}$s$^{-1}$.

\subsection{Six-Day Evolution of Vertical Energy and Helicity Fluxes}
\label{temporal}

To quantify the long-term evolution of the energy and helicity fluxes
in NOAA 11158, we integrate the photospheric Poynting and helicity flux
maps over the AR's field of view ($665\times645$-pixels) and analyze
their behavior over 6 days ($768$ time steps).

\subsubsection{Evolution of Free, Potential, and Total Energy Fluxes}

 \begin{figure*}[htb!]
  \centering 
    \resizebox{0.48\hsize}{!}{\includegraphics[angle=0]{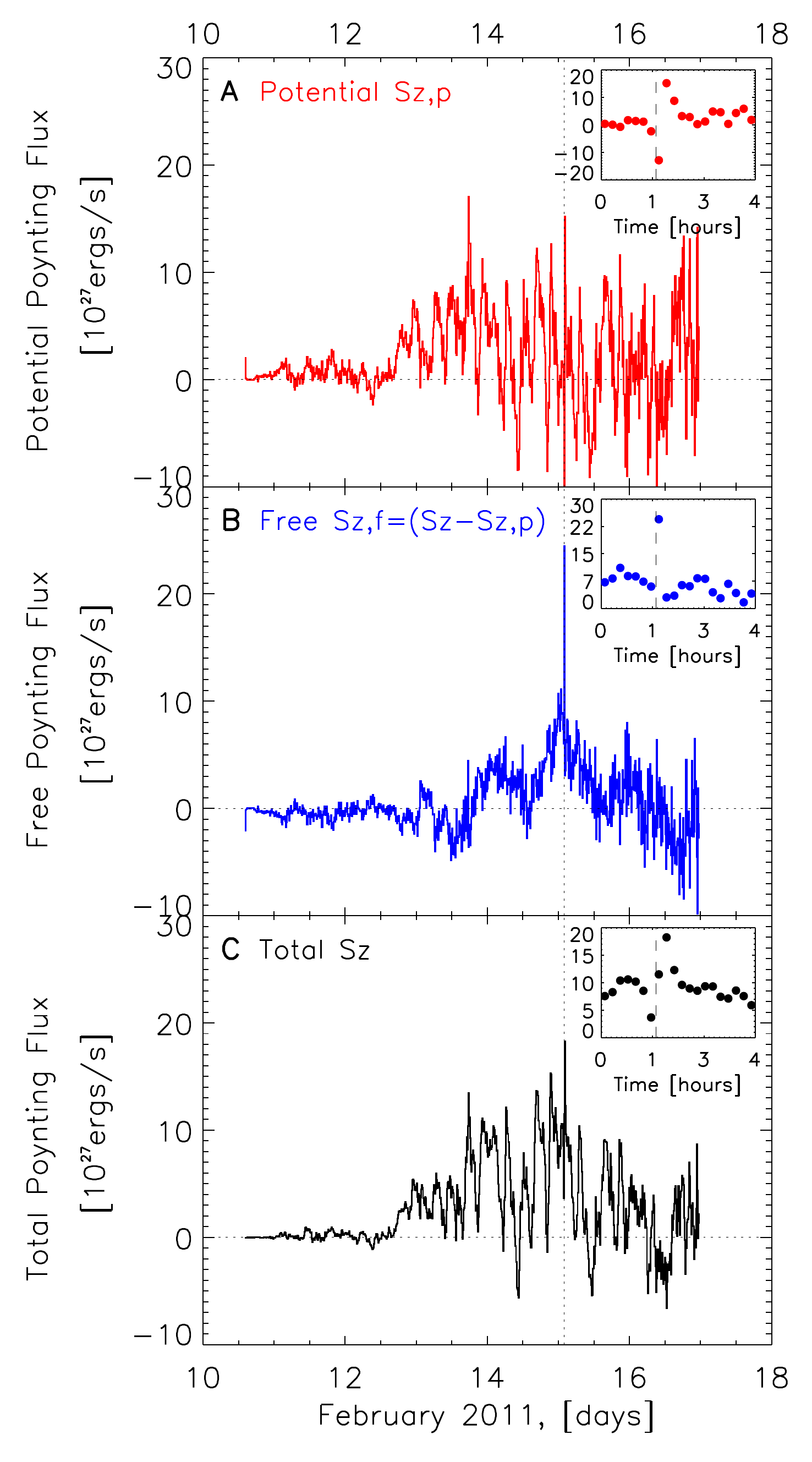}}
    \resizebox{0.48\hsize}{!}{\includegraphics[angle=0]{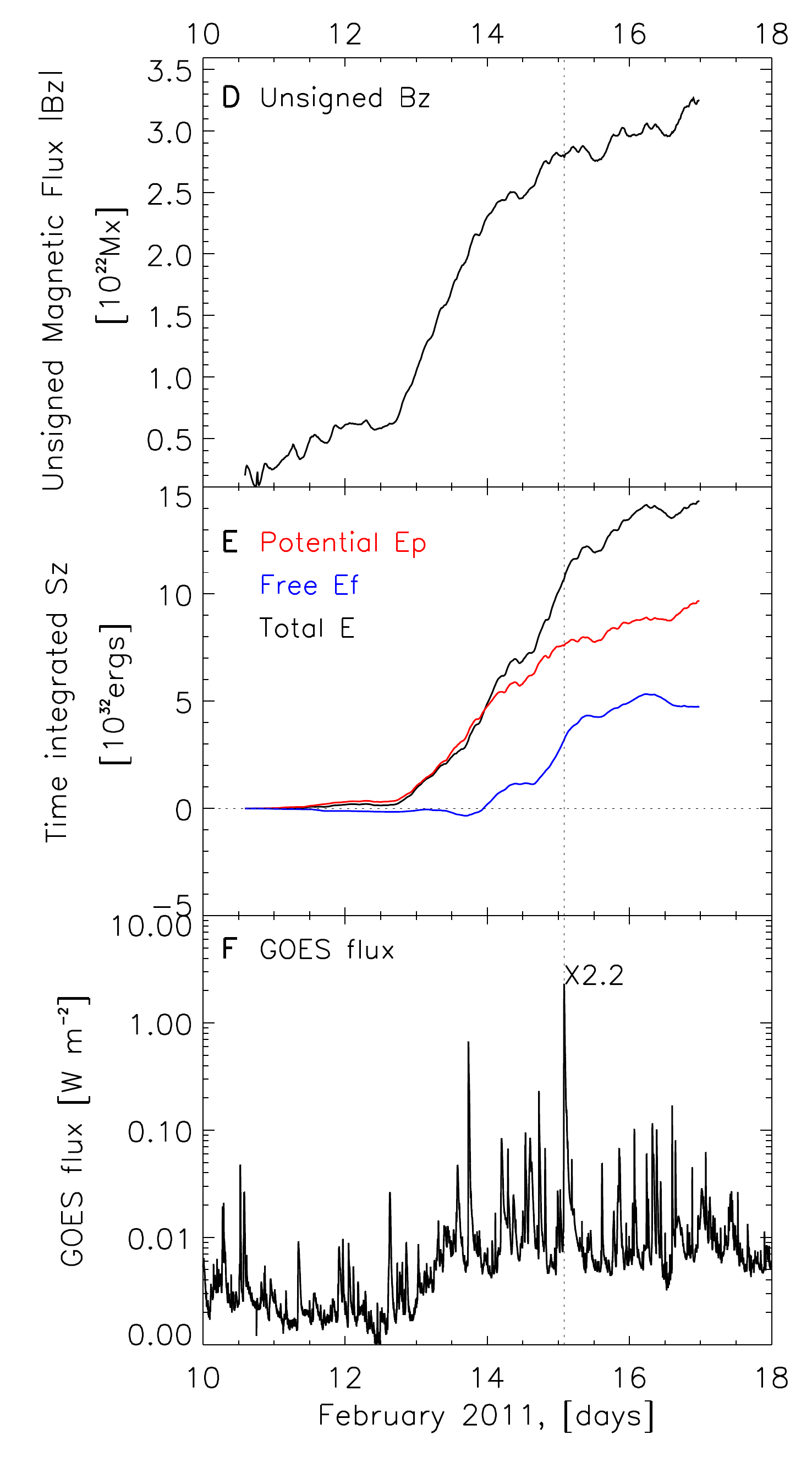}}
  \caption{ Evolution of magnetic and Poynting fluxes in NOAA 11158
    during six days: (A-C) Area-integrated potential (A), free (B) and
    total (C) Poynting fluxes. Corner insets show evolution of the
    same quantities one hour before and 2 hours after the flare; the
    X-axis is in hours; (D) Area integrated unsigned magnetic flux,
    (E) Area and time integrated free and potential components of the
    vertical Poynting flux and their sum, (F) GOES 5-minutes soft X-ray light
    curve (1-8 \AA{} channel).  The vertical dotted line indicates the
    GOES flare peak time at 01:56 UT. All the quantities were
    calculated within the FOV shown in A-D Panels in Figure~\ref{fig_mag}.}
  \label{fig_dszdt} 
\end{figure*}

Panels (A-F) of Figure \ref{fig_dszdt} show the six-day time evolution
of area-integrated potential and free vertical energy fluxes and their sum, the vertical Poynting flux: $[S_{z,p}, S_{z,f}, S_{z}]$. The potential component, $S_{z,p}$ is calculated by taking a time derivative of the coronal potential energy, computed as a surface integral over the
photosphere, with the potential function computed using a Green's
function technique. The free component is the difference between $S_z$
and $S_{z,p}$: $S_{z,f}=S_{z}-S_{z,p}$. In the three insets on the left panels we also show
evolution of $S_z$ from one hour before to four hours after the flare.
In panels A and C, large fluctuations are present in \szpot~and $S_z$
(which were calculated independently), with timescales ranging from
about $4-24$ hr.
In panel B, $S_{z,f}$ does not exhibit such strong fluctuations 
at shorter timescales, which evidently cancel out in the difference 
$S_z$ - \szpot; a residual 24-hr. periodicity can still be seen, however.
These fluctuations are related to orbital motions of the SDO satellite
\citep{Liu2012a} and are present over the whole field of
view. Unfortunately, these fluctuations are internal to the HMI
polarization measurements, hence their removal is a time-consuming and
complicated task.  Efforts to address these artifacts are underway,
but they have not been fully addressed at the time this article was written. Fortunately, when these
fluctuations are integrated over several hours, they do not cause
large fluctuations in the total cumulative fluxes (see panel E).


%
How do \szpot, $S_{z,f}$, and $S_z$ change during the flare?  Insets of Panels A-C of Figure \ref{fig_dszdt} show that $S_{z,f}$ and $S_z$ increase during the flare, after which they return to nearly preflare values.  We can understand this change in the following way. Since the change in the magnetic field close to the PIL is practically a step function \citep{Petrie2012}, the value of the Poynting flux, which involves the temporal derivative of magnetic field, should indeed appear like a delta function in time.  We conclude that while the transients  in the vertical Poynting flux include spurious signals due to flare-induced effects on the HMI spectral line \citep{Maurya2012}, the observed peak in the Poynting flux around the flare time is real. (The spatial distributions of the Poynting flux before and after the flare are shown in the bottom row of Figure~\ref{fig_eszanal}.) 
%
%




Panel D shows a steady growth in the region's total unsigned flux
after February 14, after the bulk of the region's flux has emerged.  The
growth at the time of the flare is consistent with the steady upflows
seen along the main PIL, which we expect to carry magnetic fields upward
from the solar interior, across the photosphere. A slight 24-hr
periodicity is discernible from 14 February onward.

As shown in panel E, if we integrate ~\szfree,~\szpot~ and~\sztot~in time, the fluctuations seen in panels A-C largely disappear.  Notably, no contribution from the flare transient is obvious.  We also observe that, for a time, before February 14 12:00 UT, the active region appears to possess negative free energy, a spurious result.  A probable explanation for this is that our Poynting flux estimates
are affected by the orbital motions, while the potential energy that depends only on $B_z$ suffers differently from this systematic error.  An alternative approach that might ameliorate the negative free energy budget is modification of boundary conditions  used to compute the potential field -- if instead of the Neumann boundary condition, given by $B_z$, we had used a ``hybrid'' Dirichlet-Neumann boundary condition \citep{Welsch2015}, which uses both $(\nabla_h \cdot \bvec_h)$ and $B_z$ from the HMI data, our estimate of the potential energy would be $E_{p,hybrid}=4.2\times10^{32}$ erg by the flare time (cf. $E_{p}=8.6\times10^{32}$ erg ) implying three times more free energy $E_{f,hybrid}=6.4\times10^{32}$ erg. 

To conclude, taking into account the $29\%$ uncertainty in $S_z$, due to uncertainties in the HMI data and the \texttt{PDFI} method (see Section~\ref{errors}), we find the total energy that entered the photosphere by the flare time to be $E=[10.6\pm3.1]\times10^{32}$ erg, which consists of $E_{p}=8.6\times10^{32}$ erg of potential energy and $E_{f}=2.0\times10^{32}$ erg of free energy.

\subsubsection{Helicity Flux Evolution}

 \begin{figure*}[htb!]
  \centering 
  \resizebox{1.0\hsize}{!}{\includegraphics[angle=0]{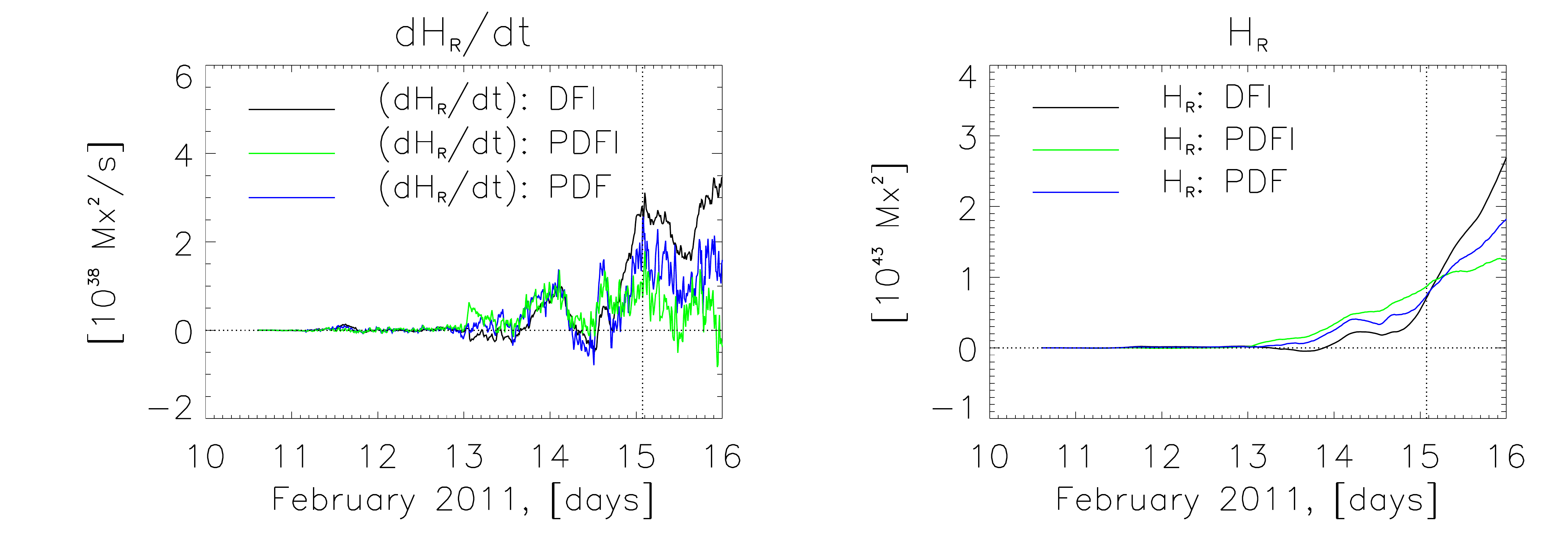}}
  \caption{ Evolution of helicity flux rates and helicity fluxes in NOAA 11158 during six days: {\it Left:} Spatially integrated helicity flux rates $\frac{dH_R}{dt}$ calculated from velocity field (\texttt{DFI}, black) and
   \texttt{PDFI} (green) and \texttt{PDF} (blue) electric fields; {\it Right}: Total helicity fluxes $H_R$, i.e. quantities on the left panel integrated in time.}
  \label{fig_hel}
\end{figure*}

Figure~\ref{fig_hel} shows time evolution of area integrated helicity
flux rates $\frac{dH_R}{dt}$ (left) and helicity fluxes $H_R$ (right) in NOAA
11158 calculated using the \texttt{PDFI} electric fields (\texttt{PDFI}) and the velocity field alone (\texttt{DFI}). The apparent large-scale 24-hour periodicity is most likely due to signal contamination from the satellite motion \citep{Liu2012a}. Until 12:00 UT on February 14 2011, the two helicity flux rates match each other, but at later times they diverge: $\frac{dH_R}{dt}$ from the velocity field (\texttt{DFI}, black) becomes significantly larger than that from the \texttt{PDFI} electric field (green).  What could be the cause of this difference? Note that the primary extra input to the \texttt{PDFI} results compared to the \texttt{DFI} results is information about $\dot{J_z}$, which sets the inductive $E_z$; this, in turn, is coupled to ${\bf E}_h$ through the ideal Ohm's constraint, ${\bf E \cdot B} = 0$.  To investigate the role that the ideal Ohm's constraint plays in the \texttt{PDFI} helicity estimate, we calculated the helicity flux rate from the non-ideal \texttt{PDF} (i.e. PTD-Doppler-FLCT) electric fields (see blue curve on Figure~\ref{fig_hel}). We find that the \texttt{PDF} and the \texttt{DFI} helicity flux rates have a similar time evolution, although the \texttt{PDF} helicity flux rate is somewhat smaller than the \texttt{DFI} rate. The discrepancy between \texttt{PDFI} and \texttt{DFI} helicity flux rates could arise because the \texttt{DFI} helicity flux rate is only sensitive to emergence and shearing. Shearing motions however, can then induce the photospheric field to unwind via the propagation of torsional Alfv\'en waves into the interior, along with the accompanying $\dot{J_z}$ and $E_z$ signatures, that only \texttt{PDFI} captures (cf., the transient twist variations modeled by \citealt{Longcope2000c}). The constraint  ${\bf E \cdot B} = 0$ then couples the resulting change in $E_z$ (and $\dot{J_z}$) to a change in $E_h$, thereby further reducing the \texttt{PDFI} helicity flux rate.  
Integrating over time we find that the total accumulated helicity flux
injected through the photosphere from the start until the flare time (right panel) is
$H_{R,DFI}=7\times10^{42}$ Mx$^2$, which is similar to the value from 
\texttt{PDFI}: $H_{R,PDFI}=8.5\times10^{42}$ Mx$^2$.


\section{DISCUSSION }\label{disc}


Here, we discuss how electric fields, energy fluxes
(Poynting fluxes), and helicity fluxes estimated with the
\texttt{PDFI} method compare to earlier results.

Using the \texttt{PDFI} method, we find photospheric electric field
components in NOAA 11158 whose amplitude varies from $-2$ V/cm to $2$ V/cm. While strong
horizontal electric fields are present in the whole active region, with the amplitude
varying from $0$ to $1.2$ V/cm, the largest concentrations of strong vertical electric field are
located mostly near the PIL and the sunspot penumbrae with the
maximum amplitudes reaching $1.5$ V/cm directed inward to the sun. As a
result of magnetic field changes during the flare, the horizontal
electric field (from the steady upward Doppler flow crossed with the
horizontal magnetic field along the PIL) increases by up to $0.5$ V/cm
perpendicular to the PIL, while the vertical electric field at PIL
remains nearly constant $E_z\approx-1.5$ V/cm.

How do our estimates of the photospheric electric fields compare to
results from earlier work? To our knowledge, there have been very few
attempts to estimate all three components of the electric field in the
photosphere.  For example, using LCT, \cite{JLiu2006} examined only
the vertical component of inductive electric field, finding the
maximum of $E_z$ to be $E_z\approx[0.1-0.2]$ V/cm, i.e. significantly
smaller than our estimates. Other attempts have focused on estimating
the electric field in the corona inside the reconnecting current sheet
(RCS) \citep{Poletto1986, Wang2003a, Wang2004a, Qiu2002, Qiu2004,
  Jing2005}. For example, using a relationship between the electric
field along the current sheet and the observable velocity and magnetic
field \citep{Priest2000}, \cite{Poletto1986} derived the maximum value
of electric field in the RCS to be $2$ V/cm. In a similar way,
\cite{Wang2003a} found the coronal electric field during the
two-ribbon flare occurring in two stages: the coronal electric field
remained near $1$ V/cm averaged over 20 minutes during the first stage, and
was followed by values of $0.1$ V/cm over the next 2 hours.
Finally, in one of the latest studies on the subject, \cite{Jing2005}
analyzed a sample of $13$ two-ribbon flares, and found the maximum electric
field inside the RCS to be in the range of $0.2-5$ V/cm.  To summarize
the above, most of the estimates of electric fields inside the reconnecting
current sheet fall within the range of photospheric electric fields
that we find here for NOAA 11158.



Using the electric and magnetic fields, we estimate the three
components of the photospheric Poynting flux vector. We find the
amplitude of the maximum Poynting fluxes in the active region to be
$S_z\approx 2\times 10^{10}$ ergs$\cdot$cm$^{-2}$ s$^{-1}$,
i.e. several times smaller than the steady state solar luminosity, $6
\times 10^{10}$ ergs$\cdot$cm$^{-2}$ s$^{-1}$, but of the same order of magnitude.  If we look at the values of $S_z$ across the AR, where most of the magnetic energy is coming from, we find even smaller values, ranging from $10^9$ to $10^{10}$ ergs$\cdot$cm$^{-2}$
s$^{-1}$. A question naturally follows this calculation: Is
such a Poynting flux consistent with the amount of magnetic energy
stored in the coronal part of the active region? This is the question
we address in Table~\ref{tab_res}.

In Table~\ref{tab_res}, we compare estimates for coronal energy and
helicity for NOAA 11158 that we derive in this paper with results from
other papers. In the first four rows, we show results from {\it Evolutionary estimates}, i.e. works by
\citet{Tarr2013, Liu2012, Tziotziou2013, Vemareddy2012b} and \citet{Jing2012}, where
energy or helicity or both are found from the evolution of photospheric
magnetic fields, by summing the energy or helicity flux rates
injected through the photosphere from the beginning of the magnetogram
sequence until the moment before the flare.  In the last five rows, we
show results from {\it Instantaneous estimates}, i.e. works by \citet{Malanushenko2014, Aschwanden2014,Sun2012,Tziotziou2013} and \citet{Jing2012}, where a single close to the flare time magnetogram or EUV image is used to estimate the energy or helicity of the corona.
The ``Method'' column shows the type of method used to find the energy
and helicity estimates on the right. In some papers, other methods
were used to calculate the potential energy; they are indicated with a
letter next to the estimate. In addition, we indicate the type of
input data used in the calculations. This helps to explain the
difference in results between some papers that used the same methods,
but applied them to different input data, e.g. differences in
estimates of the potential field energy, $E_P$, by \citet{Tarr2013}
and \citet{Sun2012}, where $B_{LOS}$ and $B_z$ have been used
respectively.


The total magnetic energy from different models, shown in
Table~\ref{tab_res}, ranges from $6\times10^{32}$ ergs 
\citep{Malanushenko2014} to $12\times10^{32}$ ergs
\citep{Tziotziou2013}. We noticed that both coronal NLFFF methods,
which use the EUV coronal loops instead of the transverse magnetic
field as a constraint for the NLFFF extrapolation
\citep{Malanushenko2014, Aschwanden2014}, derive total energies that
are the smallest of all total energy estimates: $6\times10^{32}$ ergs
and $8.6\times10^{32}$ ergs respectively. In contrast, the
photospheric NLFFF methods, which use a vector magnetogram for
extrapolation \citep{Sun2012, Tziotziou2013}, yield the largest
estimates for the total energy: $10.6\times10^{32}$ ergs and
$12\times10^{32}$ ergs respectively. The
evolutionary estimates, that derive the total energy by integrating the energy flux,
inferred from the photospheric velocity or electric fields, yield total energies in between the
two. For example, using the MCC method, \citet{Tarr2013} derive
$E=8.5\times10^{32}$ergs. Using DAVE4VM approach, \cite{Liu2012a} and
\cite{Tziotziou2013} yield similar estimates of $E=8.8\times10^{32}$
ergs and $E=8.0\times10^{32}$ ergs, respectively. 
Finally, in this paper using the
\texttt{PDFI} method we find the total energy of $E=[10.6\pm
  3.1]\times10^{32}$ ergs. To conclude, taking the \texttt{PDFI} and HMI
uncertainties into account, we find that the total energy, estimated
right before the flare, is consistent with $E$ from DAVE4VM, MCC and
NLFFF and is slightly larger than the coronal NLFFF estimates.

If we look at the temporal evolution of energy $E$, we notice that $E$
derived from \texttt{PDFI} (Figure~\ref{fig_dszdt}) is consistent
with $E$ from DAVE4VM (Figure 14 in \citet{Liu2012a}) and NLFFF
(Figure 4 in \citet{Sun2012}).  In fact, the energies $E$ derived from the
\texttt{PDFI} and DAVE4VM are almost identical until February 14
18:00, and then start diverging several hours before the flare. By the
end of magnetogram sequence, at 18:00 UT on 16 February, the $E$ from
DAVE4VM and \texttt{PDFI} are $12\times10^{32}$ ergs and
$14\times10^{32}$ ergs, respectively.  Still, this discrepancy lies
within the uncertainty of the total energy estimate  ($29\%$, see Section~\ref{errors}).  Comparing the
temporal evolution of $E$ from \texttt{PDFI} and NLFFFs, we again find
that they are consistent with each other, with less than $15\%$
differences between the two, which is within our  uncertainty ($29\%$, see Section~\ref{errors}).


The potential field energy from different models, also shown in
Table~\ref{tab_res}, ranges from $4.8\times10^{32}$ ergs
\citep{Malanushenko2014} to $8.6\times10^{32}$ ergs (this paper).
Similar to our approach, \cite{Sun2012} used the Green's function and
$B_z$ to estimate potential field energy of $8.0\times10^{32}$ ergs, that,
within the HMI uncertainty, is consistent with our estimate.
The potential field energies calculated by \cite{Tarr2013,Malanushenko2014} and \cite{Aschwanden2014} are smaller than $E_p$ from this paper and from \cite{Sun2012}. This difference might be due to the fact, that in contrast to this paper and \cite{Sun2012}, where $B_z$ is used to calculate $E_p$, \cite{Tarr2013,Malanushenko2014} and \cite{Aschwanden2014} used $B_{LOS}$ that tend to underestimate the field at $B_{LOS}>1000$ G due to limitation of the LOS pipeline algorithm (see Section 7.2.3 and Figure~17 in \citealt{Hoeksema2014}) and therefore lead to underestimate of energies by a factor of $1.4-1.7$ \citep{Liu2012a,Malanushenko2014}, that we find to be consistent with the differences in $E_p$ in Table~\ref{tab_res}.




The free magnetic energy from different models, also shown in
Table~\ref{tab_res}, ranges from $1.0\times10^{32}$ ergs
\citep{Aschwanden2014} to $2.9\times10^{32}$ ergs
\citep{Tarr2013}. Using the MCC model and the flare ribbon locations,
which allow one to derive the footprint of the reconnecting magnetic
fields, \cite{Tarr2013} find the initial pre-flare free energy in the
corona to be $E_{f}=2.9\times10^{32}$ ergs, consistent with estimates
of \cite{Sun2012} and the results of this paper. \citet{Tarr2013} also
find that more than $50\%$ of this energy, $dE=1.7\times10^{32}$ ergs,
is released during the flare. Coronal NLFFF methods
\citep{Aschwanden2014, Malanushenko2014} find that $60\%$ to
$80\%$ of the free energy is released during the flare, but the values
of the pre-flare free energy $E_f$ that they find are roughly factors
of two (or more) times smaller than the values from \citet{Tarr2013,
  Sun2012}, and this paper.


Another important fact one must keep in mind when comparing cumulative
free Poynting fluxes and coronal free energies, is that the total
Poynting flux only gives us information about the total energy that
entered the corona from the photosphere. What we do not know is how
much of this total energy leaves the corona from above into the
heliosphere during eruptive flares prior the X2.2 flare. For this
reason, our \texttt{PDFI} free energy estimate is the upper limit of
the energy available in the corona.  To summarize, the free magnetic
energy, which we find from a difference of integrated Poynting flux and the potential energy, is up to two times larger than the free energy
estimated from the coronal NLFFF codes, and $20-30\%$ smaller than
free-energy estimates from the MCC model and photospheric NLFFF
extrapolations \citep{Tarr2013, Sun2012}. The temporal evolution of the
\texttt{PDFI} free energy is within $20-30\%$ of $E_f(t)$
from the photospheric NLFFF code.



Finally, in the last column of Table~\ref{tab_res}, we compare the
total relative magnetic helicities before the X2.2 flare.  The time
integrated total helicity flux we find with \texttt{DFI} and \texttt{PDFI}
techniques is $H_{R,DFI}=7\times10^{42}Mx^2$ and
$H_{R,PDFI}=8.5\times10^{42}Mx^2$ respectively. The difference between
the two is consistent with the results we found for the ANMHD test
case, where the \texttt{DFI} method reconstructed the total helicity around $10\%$
more accurately than the \texttt{PDFI} method (see Table 3 in
\cite{Kazachenko2014a}). For comparison, using DAVE4VM,
\citet{Liu2012a} and \citet{Tziotziou2013} found that the total amount
of helicity injected into corona is $6.5\times10^{42}$ Mx$^2$ and
$8.5\times10^{42}$ Mx$^2$ respectively. These results are consistent
with our \texttt{PDFI} estimate, given the differences between the
DAVE4VM and the \texttt{PDFI} accuracies \citep{Kazachenko2014a} --
the ratio between the total \texttt{PDFI}- and DAVE4VM-reconstructed helicity
fluxes and the actual helicity flux are $0.94$ and $1.0$ respectively
for the ANMHD test case (see Table~4 in \cite{Kazachenko2014a}). Using
a different velocity reconstruction method, DAVE,
\cite{Vemareddy2012b} and \cite{Jing2012} find the total relative
helicity of $6.0\times10^{42}$ Mx$^2$ and $5.5\times10^{42}$ Mx$^2$
respectively. As shown by \cite{Schuck2008} for the ANMHD test case,
combining DAVE flows with ANMHD's vertical velocity overestimates the
total helicity flux by at least $40\%$, hence the disagreement between
DAVE and \texttt{PDFI} results for NOAA 11158 is not surprising.
Finally, using different NLFFF approaches, \citet{Tziotziou2013} and
\citet{Jing2012} find $H_R=13 \times10^{42}$ Mx$^2$ and $H_R=5.2
\times10^{42}$ Mx$^2$ respectively. To summarize, using the
\texttt{PDFI} method we find the relative magnetic helicity consistent
with DAVE4VM estimates \citep{Liu2012a, Tziotziou2013}, but very
different from (and in between) the NLFFF estimates \citep{Tziotziou2013, Jing2012}.

\begin{table*} 
\caption{Summary Table: Comparison of coronal energy and helicity in
  NOAA 11158 where an X2.2 flare occurred on February 15 2011 01:35 UT:
  $dE$ - change in the coronal free magnetic energy during the flare,
  $[E_{f},E_{p},E]$ - free, potential and total magnetic energies in
  the corona before the flare at 01:35 UT, $H_R$ -- relative magnetic
  helicity in the corona before the flare at 01:35 UT. In {\it
   Evolutionary} or Photospheric estimates, coronal energy and helicity are
  calculated cumulatively from tracking magnetic field evolution, i.e. by integrating photospheric energy and
  helicity flux rates in time. In {\it Instantaneous} or Coronal Estimates energy and
  helicity are calculated instantaneously, using extrapolation of the
  photospheric vector magnetic fields. All quantities have been
  calculated using the indicated {\it Method}, unless a more specific
  method is used, like in the case of potential energies (see letters
  $^{(d)-(f)}$).  In photospheric estimates we use a start time of
  14:11 UT on 10 February. }
  \small
 \begin{center}
 \ra{0.65}
 \begin{tabular}{@{}llllllllll@{}}\toprule
 Paper &Method &Data &  & $dE$ & $E_{f}$&$E_{p}$&$E$ &  &$H_R$ \\
 \cmidrule{5-8} \cmidrule{10-10} 
 & & & & \multicolumn{4}{c} {$10^{32}$ ergs} &  & $10^{42}$ Mx$^2$ \\ \midrule
\multicolumn{9}{l}{{\bf{Evolutionary Estimates}}}\\ 
\hspace{10pt}{\bf This paper}& {\bf \texttt{PDFI} Method}& ${\bf B},{\bf V_z}$ && -- & ${\bf 2.0}$ & ${\bf 8.6^{(d)}} $ & ${\bf 10.6}$ & &  ${\bf 8.5}$ \\
\hspace{50pt}{\it ...} & DFI method & ${\bf B},V_z$ && -- &-- & -- & -- &&  $7.8$\\
\hspace{10pt}\cite{Tarr2013}  &MCC model& $B_{LOS}$ &&$1.7$ & $2.9$ &$5.6^{(d)}$&$8.5$ &&  --  \\
\hspace{10pt}\cite{Liu2012a}  &DAVE4VM& ${\bf B},V_z$ &&-- & -- & -- & $8.8$ & &6.5  \\
\hspace{10pt}\citet{Tziotziou2013} & DAVE4VM & ${\bf B},V_z$ && -- & -- &-- &$8.0$ & &$8.5$  \\
\hspace{10pt}\citet{Vemareddy2012b} & DAVE & ${\bf B}$  && --  & -- &-- &-- & &$6.0$  \\
\hspace{10pt}\cite{Jing2012}  &DAVE& ${\bf B}$  &&-- & -- & --& -- && $5.5$   \\
\multicolumn{9}{l}{{\bf{Instantaneous Estimates}}}\\ 
\hspace{10pt}\cite{Malanushenko2014} & Coronal NLFFF$^{(a)}$& $B_{LOS}$ &&$1$ &$1.2$ &$4.8^{(e)}$ &$6$ &&  --  \\
\hspace{10pt}\citet{Aschwanden2014}  & Coronal NLFFF$^{(a)}$& $B_{LOS}$ && $0.6$ & $1.0$ &$7.6^{(f)}$ & $8.6$& & -- \\
\hspace{10pt}\cite{Sun2012} & NLFFF method$^{(b)}$& ${\bf B}$  && $0.3$ &$2.6$ &$8.0^{(d)}$ &$10.6$ &&  --  \\
\hspace{10pt}\citet{Tziotziou2013} & NLFFF  method$^{(c)}$& ${\bf B}$  &&$0.8$ & -- &-- &$12$ & &13  \\
\hspace{10pt}\cite{Jing2012}  &NLFFF method$^{(b)}$& ${\bf B}$  &&-- & -- & --& -- && $5.2$   \\
\bottomrule
\multicolumn{9}{l}{\footnotesize{
{\it NLFFF method}:  
$^{(a)}$ -- EUV loops instead of $B_t$ are used as a constraint, 
$^{(b)}$ -- \cite{Wiegelmann2004},
}}\\ 
\multicolumn{9}{l}{\footnotesize{
$^{(c)}$ -- \cite{Georgulis2012}}.
{\it Potential field methods:}
\footnotesize{
 $^{(d)}$ -- Green's function, \cite{Sakurai1989},
}}\\ 
\multicolumn{9}{l}{\footnotesize{
$^{(e)}$ -- Green's function, \cite{Chiu1977}, 
$^{(f)}$ -- \cite{Aschwanden2010}.
}}\\ 

\end{tabular}
 \normalsize \label{tab_res} \end{center} \end{table*}

\section{CONCLUSION }\label{conc}

The electric field on the Sun plays an important role in transporting
energy, heating plasma, and accelerating and transporting charged
particles. Estimates of photospheric electric and magnetic field
vectors make the estimation of Poynting flux of electromagnetic
energy crossing the photosphere and the flux of relative magnetic
helicity straightforward. Taking advantage of the newly released high
temporal and spatial resolution HMI vector magnetograms \citep{Schou2012}, and the
recently developed \texttt{PDFI} electric-field inversion method
\citep{Kazachenko2014a}, we apply the method to a six-day sequence of
vector magnetic field measurements of NOAA 11158, from 10 to 16 February
2011.  From these 12-minute cadence measurements, we derive the evolution of
electric fields, the Poynting flux, and the helicity flux.  During the
interval of study, an X2.2 flare occurred, along with 35 M- and
C-class flares.

We analyze the spatial distribution of the derived electric field and
Poynting flux maps, their temporal evolution, and changes during the
X2.2 flare. We compare derived \texttt{PDFI} electric fields with various estimates of a typical coronal and photospheric electric fields made to date, \texttt{PDFI} energies and helicities with those previously reported in the literature for this active region (NLFFF, MCC and DAVE4VM estimates). The results are the following:

\begin{enumerate}

\item We find the photospheric electric field vector, which
  typically ranges from $-2$ to $2$ V/cm, to increase its magnitude by
  up to $0.5$ V/cm at the PIL and $1$ V/cm  away from the PIL during the flare. The horizontal component is mostly concentrated along the PIL, while the vertical component is largest
  at the PIL and in the sunspots' penumbrae. The range of photospheric
  electric fields is consistent with the coronal electric fields,
  derived from the motion of flare ribbons.

\item We find the photospheric Poynting flux ranging from $[-0.6$ to
  $2]\times10^{10}$ erg$\cdot$cm$^{-2}$s$^{-1}$ with majority of the energy flux moving upward into corona and more than half of the total energy input rate injected from within the range of $[10^9$ to $10^{10}]$ erg$\cdot$cm$^{-2}$s$^{-1}$. The largest
  vertical Poynting flux is concentrated at the PIL. 

 \item Integrating the Poynting flux in time we find the total magnetic
   energy before the flare, $E=[10.6\pm3.1]\times10^{32}$ ergs. In spite of a
   very different approach, it is consistent within the uncertainty
   with the total energies from DAVE4VM, MCC and NLFFF methods'
   estimates and larger than the coronal NLFFF estimates.

 \item The potential field magnetic energy before the flare, estimated via the Green's function, $E_p=8.6\times10^{32}$ ergs, is consistent with $E_p=8.0\times10^{32}$ergs
   \citep{Sun2012}. \citet{Tarr2013, Malanushenko2014} and \citet{Aschwanden2014}
   derive a similar or somewhat smaller $E_p$, using different computation methods and $B_{LOS}$ instead of
   $B_z$ that is prone to being underestimated in the strong field regions due to limitation of the LOS pipeline algorithm \citep{Hoeksema2014}.
   
  \item The free magnetic energy before the flare from the \texttt{PDFI} method,
    $E_f=2.0\times10^{32}$ ergs, is up to two times larger than the
    free energy estimated by coronal NLFFF codes \citep{Aschwanden2014,Malanushenko2014}, and $20-30\%$ smaller
    than the free energy estimates from MCC model and photospheric
    NLFFF extrapolations \citep{Tarr2013,Sun2012}.

  \item Analyzing the temporal evolution of cumulative energy, $E$,
    from \texttt{PDFI}, NLFFFs and DAVE4VM, we find less than $15\%$ differences
    between the three. The temporal evolution of the \texttt{PDFI}
    $E_f$ is less than $20-30\%$ different from the $E_f$ from the
    photospheric NLFFF codes and several times larger than $E_f$ from the coronal NLFFF codes \citep{Aschwanden2014, Malanushenko2014}.
  
  \item We find the relative magnetic helicity to be consistent with
    DAVE4VM estimates \citep{Liu2012a, Tziotziou2013}, but very
    different from the NLFFF estimates \citep{Tziotziou2013,
      Jing2012}.


\item Using Monte-Carlo simulations, we find that the levels of the
  errors in the HMI data lead to uncertainties in the horizontal
  electric field of  $13\%$ to $18\%$ that result in errors in the
  vertical Poynting flux of around $14\%$. If we add those
  uncertainties to the errors that we found when testing the
  \texttt{PDFI} method ($<25\%$ in $S_z$, \citet{Kazachenko2014a}),
  then we end up with $14\%$ to $29\%$ errors in the vertical Poynting
  flux depending on the LOS angle. Also, since some free energy might
  have been released by flares prior to the X2.2 flare, we view the
  \texttt{PDFI} estimates of the free energy injected through the
  photosphere to be upper limits on the total free energy available
  before the flare.
  \end{enumerate}
  
This study is the first 
application of the \texttt{PDFI}
electric field inversion technique to photospheric vector magnetic field
and Doppler measurements.  We find that the total amount of energy and
helicity injected through the photosphere before the flare estimated
by the \texttt{PDFI} method is consistent with estimates from other
approaches, in spite of differing techniques. This agreement is
very promising, implying that the \texttt{PDFI} technique is not only
capable of describing the coronal energy and helicity budget, but can also
provide instantaneous estimates of energy and helicity transferred
through the photosphere. 

We believe that both the derived dataset of \texttt{PDFI} electric
fields and the \texttt{PDFI} method itself will be useful to the
science community for analysis of the evolution and spatial
distribution of the photospheric electric fields, fluxes of energy and
helicity, and their relationships with flare activity.  In addition,
\texttt{PDFI} electric fields can be used as time-dependent boundary
conditions for data-driven models of coronal magnetic field
evolution \citep{Fisher2015a}. The dataset for NOAA 11158 is available for downloading on
our website\footnote{\url{http://cgem.stanford.edu}}.


\acknowledgments
\acknowledgments
We thank the US taxpayers for providing the funding that made this research possible. We thank the anonymous referee for thoughtful input that have improved
the manuscript. We acknowledge funding from the Coronal Global Evolutionary Model (CGEM) award NSF AGS 1321474 (MDK, BTW, GHF), Coronal Global Evolutionary Model (CGEM) award  NASA Award NNX13AK39G (XS, YL),  NSF Award AGS-1048318 (GHF), NASA Award NNX13AK54G (MDK), NSF SHINE Postdoc Award 1027296 (MDK),  NSF's National Space Weather Program AGS-1024862 (BTW), the NASA Living-With-a-Star TR\&T Program NNX11AQ56G (MDK, BTW, GHF), and the NASA Heliophysics Theory Program NNX11AJ65G (GHF,  BTW).
\appendix
\section{SCALING  {\bvec} AND {\evec} TO CARTESIAN MERCATOR MESH}\label{merc}

The data we analyze have been transformed from plane-of-sky to Mercator projection
with a local, Cartesian coordinate system centered on NOAA 11158.  The
distortion of pixel scales in this transformation has implications for
inferring electric fields from magnetic evolution that we describe in
detail here.

When observed magnetic fields are reprojected from the Sun's observed plane-of-sky surface onto a plane using a conformal (angle-preserving) mapping,
pixel areas in different regions of the new coordinate system
generally do not correspond to the same physical areas on the
Sun. Denoting the area of a pixel $ij$ on the Sun as $A^S_{ij}$, and in
the mapped coordinates as $A^M_{ij}$, $A^M_{ij} \approx A^S_{ij}$ near
the center of the projection, but $A^M_{ij} \neq A^S_{ij}$ far from
center. In contrast, the reprojection does not directly alter the
magnetic field values themselves: the fields are interpolated onto the
new grid point, but the interpolation attempts to faithfully
represent measured field values at each point. While
observed magnetic field vector data have units of field strength,
i.e. Gauss, they are more accurately described as measurements of
pixel-averaged flux densities, Mx/cm$^2$. This is because the flux in
pixel $ij$ could be confined within a subregion of the pixel with
fraction area $f_{ij}A ^S_{ij}$, where the fill fraction $f_{ij}$ obeys
$0<f_{ij}<1$, implying a true field strength $B^{true}=B^{app}/f$
larger that the apparent field strength (``pixel-averaged flux
density'') $B^{app}$ measured by the instrument.

The question, however, arises: Do we need to modify the field
strengths in the new projection -- call the original field
strengths ${\bf B}^S$, and the reprojected field strengths ${\bf B}^M$
-- to account for the distorted areas? If so, how? Do vertical and horizontal magnetic field components need to be
compensated in the same way? How should velocities and electric fields
be modified to compensate for this distortion?

Below we consider the transformation used in this paper, from the plane-of-sky to Mercator projection
\citep{Welsch2009}. 
Briefly, the distortion of scale in cylindrical projections such as Mercator is a function of latitude alone. The horizontal
Mercator coordinate, $x$, is mapped one-to-one with the heliocentric
longitude, $\phi$, and is independent of latitude. Because the
distance between lines of constant longitude decreases with increasing
latitude, the projection's scale (the distance of the sphere corresponding
to a fixed distance in the projected image) must decrease (fewer solar
Mm per Mercator pixel) with increasing latitude. As lines of constant
longitude converge towards the poles on a spherical surface, the
physical distance between such lines goes to zero. In Mercator
coordinates, however, the distance between two lines of constant
longitude $\Delta x$, is fixed and is independent of latitude
$\theta$. Effectively, this means the projection magnifies distances
towards the poles. (This effect is easily seen on Mercator projections
of the Earth's surface, on which Antarctica and Greenland, for
example, appear too large relative to landmasses at lower latitudes.)
Consequently, displacements in the vertical
Mercator coordinate, $dy$, corresponding to a fixed latitudinal
displacement $d \theta$, should increase towards the poles. Because the
physical length corresponding to a fixed $dy$ shrinks towards the poles as
$1/cos (\theta)$, and $dx$ scales in the same way,
pixel areas in the reprojected system scale as $cos^2 (\theta)$
compared to areas on the Sun.

Consider two reprojected pixels with the same normal magnetic field,
one at a high latitude and one at a low latitude, denoting these
$B^H_n=B^L_n$ respectively. These have the same fluxes in the
projection, $\Phi^{M,H} \approx \Phi^{M,L}$. The flux in each
reprojected pixel, however, corresponds to different fluxes on the
Sun: $\Phi^{M,L} \approx \Phi^{S,L}$, but $\Phi^{S,H} < \Phi^{M,H}$
since the actual solar area corresponding to the high-latitude pixels is
smaller. We choose to handle this by multiplying $B_n$ in each
reprojected pixel by $cos^2(\theta)$, to compensate for distortion of
its area as a function of latitude; at the same time, we do not
rescale pixel lengths.

When flux emerges into a pixel, $v_n$ transports horizontal field
$B_h$ along a pixel edge of length $L$ over a time interval $\Delta t$,
meaning $\Phi_{em}=v_n L B_h \Delta t$ has emerged (see discussion in Section 2, and Figure~3 of  \cite{Welsch2013}). This means that the
normal flux in the pixels that share this edge must change by
$\Phi_{em}$, to account for the emerged flux. We have assumed that
flux in each pixel sharing L has been compensated by a factor of
$cos^2(\theta)$. For the emerged horizontal flux to match the changes
in the vertical flux, two factors of $cos(\theta)$ must be present in the
product $v_n B_h$, since we are not rescaling $L$. We can scale $v_n$
by $cos^{\alpha}(\theta)$ and $B_h$ by $cos^{\beta}(\theta)$, which
yields

\begin{equation}\label{eq_sc1}
\alpha+\beta=2.
\end{equation}
When flux is horizontally transported from one pixel to another, $v_h$
transports the vertical field $B_n$ across a pixel edge of length $L$ over
a time interval $\Delta t$, meaning a flux $\Phi_{xport}=v_h L B_n
\Delta t$ has been moved. This means the normal fluxes in the pixels
that share this edge must each change by $\Phi_{xport}$, to account for
the transported flux. Because $B_n$ has already been rescaled by
$cos^2{\theta}$, $v_h$ in the product $v_h L B_n$ does not need any
scaling.

The pixel-integrated vertical Poynting flux $S_z=(E_h \times B_h)$  at high latitude must be
scaled by $cos^2(\theta)$, to account for the area distortion. Recall
that $c E_h$ is proportional to the sum of $v_n B_h$, scaled by
$cos^{\alpha + \beta}(\theta)$, and $v_h B_n$, are already scaled by
$cos^2(\theta)$. Because $S_z$ must scale as $cos^2(\theta)$ for any
$v$, we can consider the special case $v_h = 0$, implying $S_z=v_n
B^2_h$, so

\begin{equation}\label{eq_sc2}
\alpha+2\beta=2.
\end{equation}

Comparing equations~\ref{eq_sc1} and~\ref{eq_sc2} yields $\beta=0$,
so $B_h$ is not changed, but $\alpha = 2 $, so $v_n$ is scaled by
$cos^2(\theta)$.  Physically, this implies that the vertical transport
of emerging flux at high latitudes is scaled to make sure a ``scaled
amount'' of flux emerges. This scaling implies that $E_h$ is
automatically and implicitly scaled (via scaling applied to $B_h$ and
$v_n$) by $cos^2(\theta)$, and that $E_z=v_h \times B_h$ is unscaled.

Summarizing the above, we scale $B_n, v_n, E_h$ by $cos^2(\theta)$ and
do not scale $B_h,v_h, E_n$.




\end{document}